\documentclass[a4paper,twocolumn,10pt]{article}
\usepackage[left=1.5cm,top=3cm,right=1.5cm,bottom=3cm]{geometry} 
\usepackage{multicol}
\usepackage[latin1]{inputenc}
\usepackage{graphicx}
\usepackage{natbib}
\usepackage{caption}
\usepackage{wasysym}
\usepackage{tikz}
\usepackage{hyperref}
\usepackage{cite}

\begin{document}

\title{Spectroscopic and photometric study of the eclipsing interacting binary V495 Centauri}
\author{
Rosales, J.A.$^{1}$, Mennickent, R.E.$^{1}$, Djura\v{s}evi\'{c}, G.$^{2,3}$, Araya, I.$^{4,5}$, Curé, M.$^{4}$\\ \\
\small{$^{1}$ \emph{Departamento de Astronomía,Universidad de Concepción,Casilla 160-C, Concepción, Chile.}}\\
\small{$^{2}$ \emph{Astronomical Observatory, Volgina 7, 11060 Belgrade 38, Serbia.}}\\
\small{$^{3}$  \emph{Isaac Newton Institute of Chile, Yugoslavia Branch.}}\\
\small{$^{4}$ \emph{Instituto de Física y Astronomía, Facultad de Ciencias, Universidad de Valparaíso, Chile.}}\\
\small{$^{5}$ \emph{Núcleo de Matemáticas, Física y Estadística, Facultad de Ciencias, Universidad Mayor, Chile.}}\\
}
\date{}
\vskip 2cm

\twocolumn[
\begin{@twocolumnfalse}
	\maketitle
\begin{abstract}

Double Periodic Variables (DPV) are among the new enigmas of semi-detached eclipsing binaries. These are intermediate-mass binaries characterized by a long photometric period lasting on average 33 times the orbital period. We present a spectroscopic and photometric study of the DPV V495 Cen based on new high-resolution spectra and the ASAS V-band light curve. We have determined an improved orbital period of $33.492 \pm 0.002$ d and a long period of 1283 d. We find a cool evolved star of $M_{2}=0.91\pm 0.2 M_{\astrosun}$, $T_{2}= 6000\pm 250 K$ and $R_{2}=19.3 \pm 0.5 R_{\astrosun}$ and a hot companion of $M_{1}= 5.76\pm 0.3 M_{\astrosun}$, $T_{1}=16960\pm 400 K$ and $R=4.5\pm0.2 R_{\astrosun}$. The mid-type B dwarf is surrounded by a concave and geometrically thick disc, of radial extension  $R_{d}= 40.2\pm 1.3 R_{\astrosun}$ contributing $\sim$ 11 percent to the total luminosity of the system at the V band. The system is seen under inclination $84.\!\!^{\circ}8$ $\pm$ $0.\!\!^{\circ}6$ and it is at a distance $d= 2092 \pm 104.6$ pc. The light curve analysis suggests that the mass transfer stream impacts the external edge of the disc forming a hot region  11\% hotter than the surrounding disc material.  The persistent $V<R$ asymmetry of the H$\alpha$ emission suggests the presence of a wind and the detection of a secondary absorption component in He\,I lines indicates a possible wind origin
in the hotspot region. \\

	\end{abstract}
\end{@twocolumnfalse}

]

%%%%%%%%%%%%%%%%%%%%%%%%%%%%%%%%%%%%%%%%%%%%%%%%%%%%%%%%%%%%%%%%%%%%%%%%%%%%%%%%%%%%%%%%%%%%%%%%%%%%%%%%%%%%%%%%%%%%%%%%%%%%%%%%%%%%%%%
%%%%%%%%%%%%%%%%%%%%%%%%%%%%%%%%%%%%%%%%%%%%%%%%%%%%%%%%%%%%%%%%%%%%%%%%%%%%%%%%%%%%%%%%%%%%%%%%%%%%%%%%%%%%%%%%%%%%%%%%%%%%%%%%%%%%%%%
%%%%%%%%%%%%%%%%%%%%%%%%%%%%%%%%%%%%%%%%%%%%%%%%%%%%%%%%%%%%%%%%%%%%%%%%%%%%%%%%%%%%%%%%%%%%%%%%%%%%%%%%%%%%%%%%%%%%%%%%%%%%%%%%%%%%%%%
{\bf 1 INTRODUCTION}\\

Double Periodic Variables are interacting binary stars of intermediate mass that show two closely linked photometric variations being the long period roughly  33 times longer than  the orbital period (\citet{2003A&A...399L..47M,2016MNRAS.461.1674M,2017SerAJ.194....1M},\citet{2010AcA....60..179P}). The nature of the second period is unknown but suspected to reflect the strength variations of a wind generated in the stream-disc impact region (\citet{2012IAUS..282..317M,2016MNRAS.461.1674M}, \citet{2008A&A...487.1129V}).   Also, DPVs are considered as one specific evolutionary step for more massive Algols, one possibly involving mild mass transfer and systemic mass loss (\citet{2008MNRAS.389.1605M}).  An interesting property of these objects is the surprising constancy of their orbital periods, which is not expected in Algols undergoing RLOF mass transfer (\citet{2013MNRAS.428.1594G}). Also the DPVs seem to be hotter and more massive than classical Algols and seem to have always a B-type component;  their orbital periods typically run between 4 and 100 days. DPVs have been found in the Galaxy (MW), Large Magellanic Cloud (LMC) and the Small Magellanic Cloud (SMC). The components of DPVs are denoted by the suffixes \emph{g} and \emph{d} appended to the parameter designation, \emph{g} denoting the primary component (C1) or gainer (hot) and \emph{d} the secondary component (C2) or donor (cold), respectively. Recently, a mechanism based on cycles of a magnetic dynamo in the donor star, the so called Applegate mechanism, was proposed as an explanation for the DPV long  cycles consistent with the almost constancy of the orbital period (\citet{2017A&A...602A.109S}).

The importance of studying DPVs is the fact that they are interesting astrophysical objects for the study of mechanisms of mass transfer in semidetached binaries, they share similarities with some Algol systems, and can potentially provide us a wealth of information on the density, tidal friction, wind processes and stellar dynamos in massive close binaries (\citet{2015A&A...577A..55D},\citet{2014ApJ...782....7D}). Hitherto only a couple of DPVs, less than 5\%, have been studied spectroscopically (e.g \citet{2008MNRAS.389.1605M}; \citet{2012MNRAS.421..862M}a,b;\citet{2013A&A...552A..63B,2014A&A...567A.140B};\citet{2013MNRAS.428.1594G};\citet{2015IAUS..307..125M,2016MNRAS.461.1674M}).

The eclipsing interacting binary V495 Cen (ASAS ID 130135-5605.5, $\alpha_{2000}$= 13:01:35.0, $\delta_{2000}$= -56:05:30.0, V= 9.95 mag, B-V= 0.52 mag, spectral type Be)\footnote{http://simbad.u-strasbg.fr/simbad} was discovered as a DPV by \citet{2014IBVS.6116....1M}, who find that the system is characterized by a long photometric cycle of 1283 d, resulting in the system with the longest period among Galactic DPVs. V\,495\,Cen has been poorly studied and shows an orbital period of 33.48177 d in the ASAS\footnote{http://www.astrouw.edu.pl/asas/} catalogue (\citet{1997AcA....47..467P}). Our study will contribute to the overall understanding of DPVs, especially those of long orbital period.
 
 In Section 2 we present a photometric analysis of V495 Cen. In Section 3 we give a brief review of the  methods of data acquisition and data reductions. In Section 4 we present the orbital and physical parameters of the system. In Section 5 we model the light curve with a special code including the light contribution of both stars and the accretion disc, and get  luminosities, radii, temperatures, surface gravities, masses and the system's inclination.  
In Section 6  we present an analysis yielding the reddening and distance of the system. Finally,  the main results of our research are summarized in Section 7. 
\\

%%%%%%%%%%%%%%%%%%%%%%%%%%%%%%%%%%%%%%%%%%%%%%%%%%%%%%%%%%%%%%%%%%%%%%%%%%%%%%%%%%%%%%%%%%%%%%%%%%%%%%%%%%%%%%%%%%%%%%%%%%%%%%%%%%%%%%%
%%%%%%%%%%%%%%%%%%%%%%%%%%%%%%%%%%%%%%%%%%%%%%%%%%%%%%%%%%%%%%%%%%%%%%%%%%%%%%%%%%%%%%%%%%%%%%%%%%%%%%%%%%%%%%%%%%%%%%%%%%%%%%%%%%%%%%%
%%%%%%%%%%%%%%%%%%%%%%%%%%%%%%%%%%%%%%%%%%%%%%%%%%%%%%%%%%%%%%%%%%%%%%%%%%%%%%%%%%%%%%%%%%%%%%%%%%%%%%%%%%%%%%%%%%%%%%%%%%%%%%%%%%%%%%%
\vskip 0.5cm 
{\bf 2 PHOTOMETRIC EPHEMERIDES}\\

We have re-analyzed the ASAS light curve considering only those 622 better quality data points labeled as A-type by ASAS (Fig.\, 1) and rejecting outliers points, contrary to the previous analysis (\citet{2014IBVS.6116....1M}) in which were considered data points labeled as A-type and B-type, the best and normal quality respectively. The data set was analyzed by Phase Dispersion Minimization (PDM-IRAF, Stellingwerf 1978)\footnote{IRAF is distributed by National Optical Astronomy Observatories, which are operated by Association of Universities for Research in Astronomy, Inc., under cooperative agreement with the National Science Foundation}, revealing an orbital period $P_{o}=33.492 \pm 0.01613$ d (the error corresponds to the difference between the first approximation by PDM and the best value found by visual inspection of the periodogram minimum) and the epoch of minimum HJD $= 2454609.71384 \pm 0.032$ d. We have noticed that the orbital period shown in ASAS (33.4817 d) does not match the value obtained by us. This is probably due to the fact that the algorithm used by PDM to search for periods is different to that used by ASAS at the moment of carrying out the periodogram.

With the long and short periods we disentangled the light curve with a code specially designed for this purpose by Zbigniew Ko\l{}aczkowski. The code fits the data with a Fourier series consisting of fundamental frequencies and harmonics and removes them. As a result we obtain the cleaned light curve with no frequencies and two light curves for the isolated orbital and long cycles as shown in (Fig.\,2). The process reveals an orbital modulation typical of an eclipsing binary type (EB), with rounded inter-eclipse
regions revealing proximity effects, and a longer cycle characterized by a quasi-sinusoidal variability typical of Double Periodic Variables,  whose full amplitude in the $V$-band is $\sim 20 \%$ with respect to the total brightness. We determined the following ephemeris for the light curves:\\

\begin{figure}
	\begin{center}
		\includegraphics[width=8 cm,angle=0]{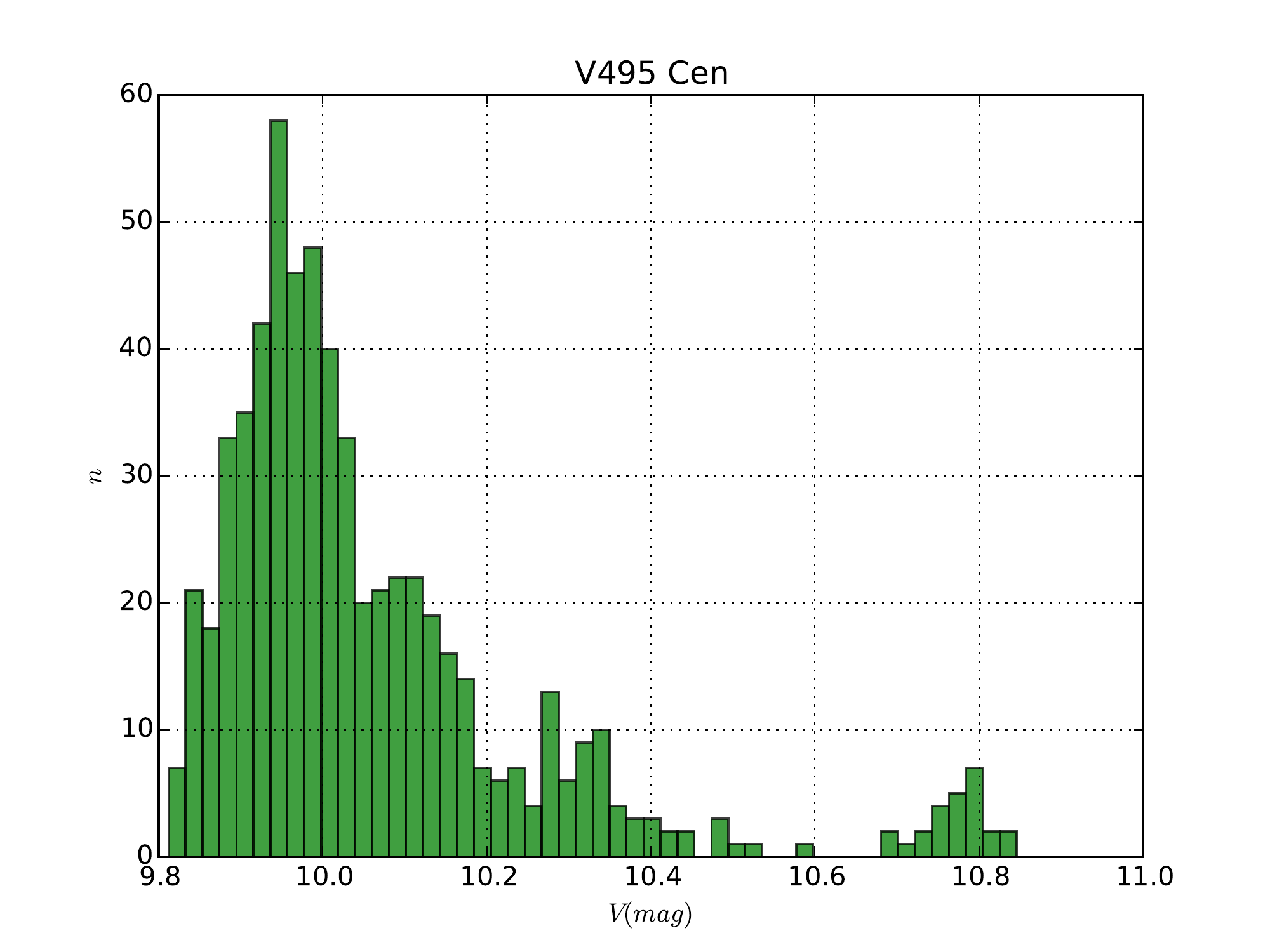}
	\end{center}
	\footnotesize
	{\bf Figure 1.} 
	Histogram of magnitude for V495 Cen with 50 bins.
	\normalsize 
\end{figure}

\begin{equation}
	HJD_{min, orbital} =2454609.713(32)+33.492(16) \times \emph{E},
\end{equation}
\\
\begin{equation}
	HJD_{max,long} = 2454926.699(42) + 1283 \times \emph{E},\hspace{1.4cm}
\end{equation}
\\
These are used for the spectroscopic analysis in the rest of the paper. \\

\begin{figure}
\begin{center}
	\includegraphics[width=8 cm,angle=0]{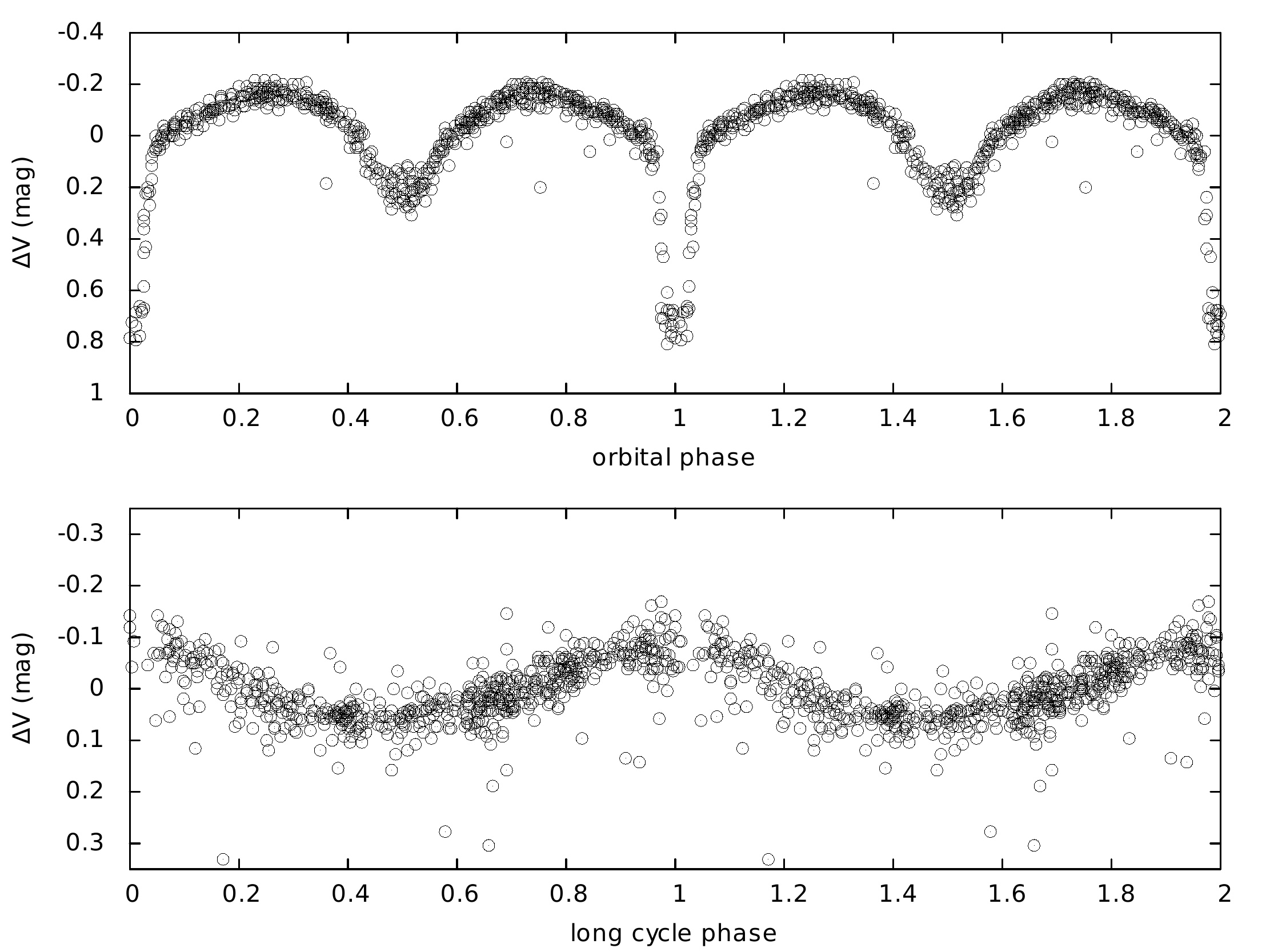}
\end{center}
\footnotesize
{\bf Figure 2.} V495 Cen ASAS V-band light curve after disentangling. Orbital period (up) and long period (down). Phases were calculated according to times of light curve minimum and maxima, given by equation (1) and (2).
\normalsize
\end{figure}

%%%%%%%%%%%%%%%%%%%%%%%%%%%%%%%%%%%%%%%%%%%%%%%%%%%%%%%%%%%%%%%%%%%%%%%%%%%%%%%%%%%%%%%%%%%%%%%%%%%%%%%%%%%%%%%%%%%%%%%%%%%%%%%%%%%%%%%
%%%%%%%%%%%%%%%%%%%%%%%%%%%%%%%%%%%%%%%%%%%%%%%%%%%%%%%%%%%%%%%%%%%%%%%%%%%%%%%%%%%%%%%%%%%%%%%%%%%%%%%%%%%%%%%%%%%%%%%%%%%%%%%%%%%%%%%
%%%%%%%%%%%%%%%%%%%%%%%%%%%%%%%%%%%%%%%%%%%%%%%%%%%%%%%%%%%%%%%%%%%%%%%%%%%%%%%%%%%%%%%%%%%%%%%%%%%%%%%%%%%%%%%%%%%%%%%%%%%%%%%%%%%%%%%
\vskip 0.5cm 
{\bf 3 SPECTROSCOPIC OBSERVATIONS}\\

To investigate the fundamental properties of V495 Cen we collected a series of high-resolution optical spectra between February and May 2015  with the CHIRON\footnote{http://www.ctio.noao.edu/noao/content/CHIRON} spectrograph. We obtained 10 spectra with $R \sim 25000$ (fiber mode) and 30 spectra with  $R \sim  80000$ (slicer mode). The spectral regions covered were 4500-8500 \AA{} and 4580-8762 \AA{} at the fiber and slicer mode, respectively, with a typical signal to noise ratio (SNR) of $\approx$ 160.

The corrections with flat and bias, wavelength calibration and order merging were done with IRAF. We also normalized all spectra to the continuum and corrected them to the heliocentric rest frame. 

The spectra obtained with CHIRON are not sky-subtracted. We have not flux calibrated our spectra but this does not affect the strength measurements and radial velocities included in this paper.\\ 
%%%%%%%%%%%%%%%%%%%%%%%%%%%%%%%%%%%%%%%%%%%%%%%%%%%%%%%%%%%%%%%%%%%%%%%%%%%%%%%%%%%%%%%%%%%%%%%%%%%%%%%%%%%%%%%%%%%%%%%%%%%%%%%%%%%%%%%
%%%%%%%%%%%%%%%%%%%%%%%%%%%%%%%%%%%%%%%%%%%%%%%%%%%%%%%%%%%%%%%%%%%%%%%%%%%%%%%%%%%%%%%%%%%%%%%%%%%%%%%%%%%%%%%%%%%%%%%%%%%%%%%%%%%%%%%
%%%%%%%%%%%%%%%%%%%%%%%%%%%%%%%%%%%%%%%%%%%%%%%%%%%%%%%%%%%%%%%%%%%%%%%%%%%%%%%%%%%%%%%%%%%%%%%%%%%%%%%%%%%%%%%%%%%%%%%%%%%%%%%%%%%%%%%
\vskip 0.7cm 
{\bf 4 SPECTROSCOPIC ANALYSIS}\\

\vskip 0.5cm
{\bf 4.1 Spectral disentangling }\\

The spectra were disentangled with a method that is quite good for separating the absorption-lines widths of both stellar components \citet{2006A&A...448..283G}. This method turns to be effective when working with absorption-lines widths less than the radial velocity amplitude of the corresponding star, i.e. for most lines except for the strong and wide H$\alpha$ emission. For this reason all spectral regions of interest  except the H$\alpha$ region were disentangled. However, as we can see in Fig.\,3, the underlying donor absorption is small compared with the strong double peak H$\alpha$ emission.  The presence of emission in H$\alpha$ confirms the interacting binary nature for this system and suggests it is in a semi-detached stage and possesses a circumstellar disc. We notice the deep central absorption between the emission peaks revealing large amounts of obscuring material through the line of sight and also the larger intensity of the red peak at this specific epoch.

\begin{figure}
	\begin{center}
		\includegraphics[width=8 cm,angle=0]{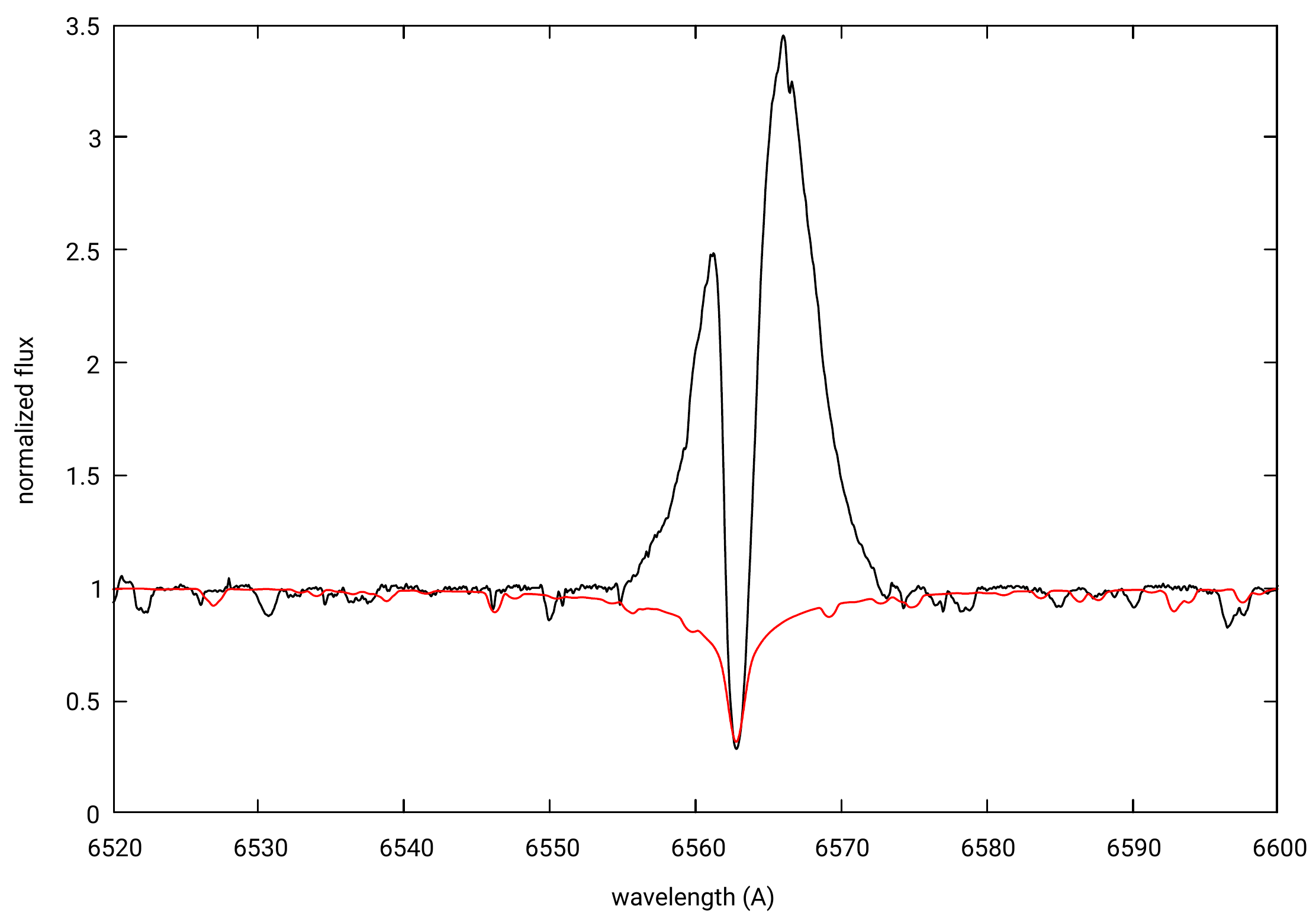}
	\end{center}
	\footnotesize
	{\bf Figure 3.} 
	The H$\alpha$ line on HJD 2457062.80354576 ($\phi_{o}=0.24$,$\phi_{l}=0.68$) and the best synthetic spectrum for the donor.
	\normalsize 
\end{figure}

\begin{figure*}
	\begin{center}
		\includegraphics[width=9 cm,angle=0]{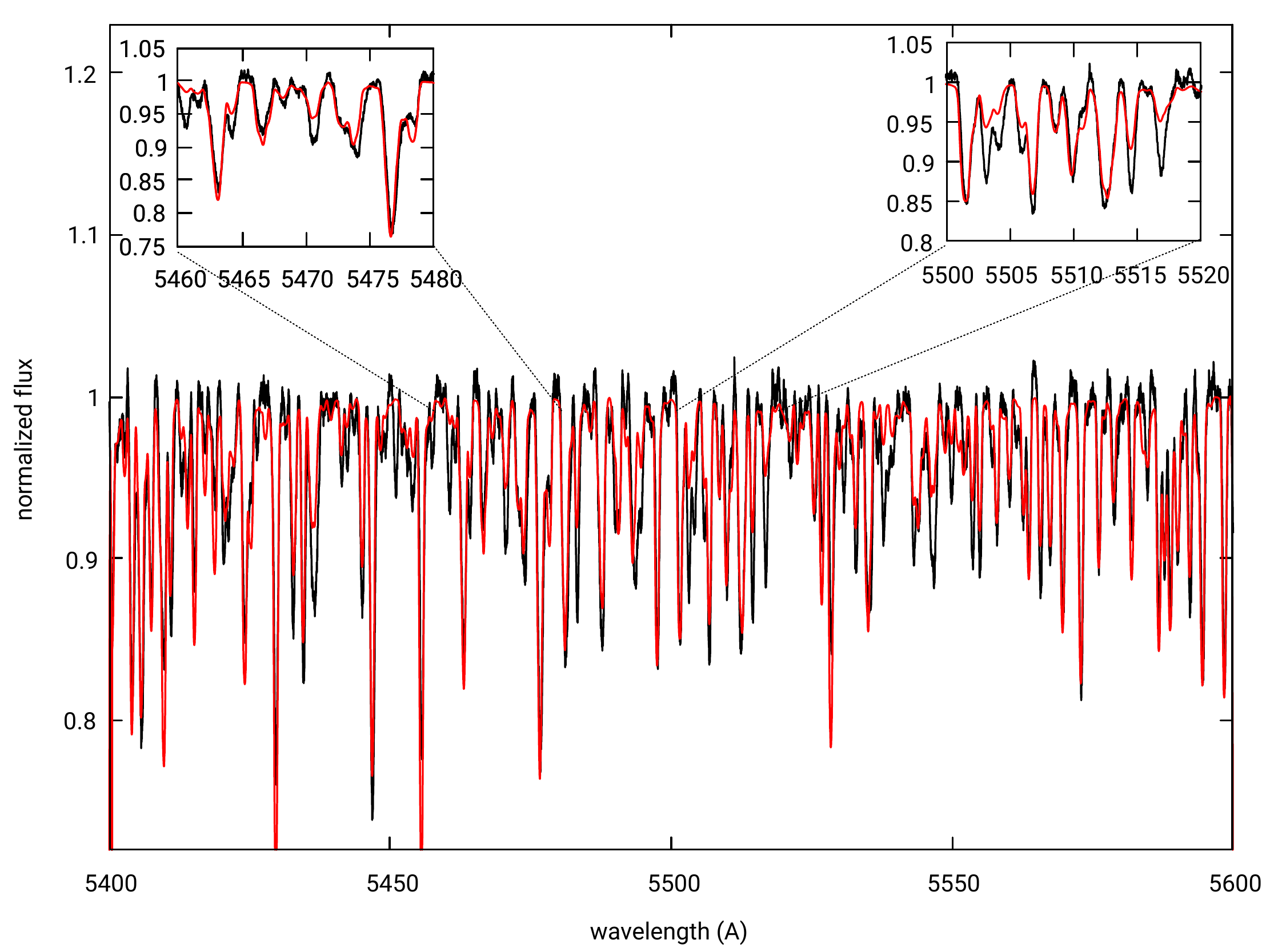}
	\end{center}
	\footnotesize
	{\bf Figure 4.} A detailed comparison between the observed and synthetic (best model: smoothed line in red colour) donor spectrum.
	\normalsize
\end{figure*}

\vskip 0.5cm 
{\bf 4.2 Determination of donor physical parameters}\\

To determine the physical parameters of the secondary star, we compared our observed average donor spectrum with synthetic spectra constructed with SPECTRUM\footnote{http://www.appstate.edu/~grayro/spectrum/spectrum.html} which uses atmospheric models computed with grids of ATLAS9\footnote{http://wwwuser.oats.inaf.it/castelli/grids.html} model atmospheres (\citet{2004astro.ph..5087C}). The models used are in Local Thermodynamic Equilibrium (LTE). 

The theoretical models were calculated for effective temperatures from 4250 to 19000 K with steps of 250 K, surface gravities from 0.5 to 3.5 dex  with steps of 0.5 dex, solar metallicity, micro-turbulent velocity from 0.0 to 2.0 km s$^{-1}$ with steps of 1.0 km s$^{-1}$ and macro-turbulent velocity from 1.0 to 20.0 km s$^{-1}$ with steps of 1.0 km s$^{-1}$. \\

\begin{figure}
	\begin{center}
		\includegraphics[width=8 cm,angle=0]{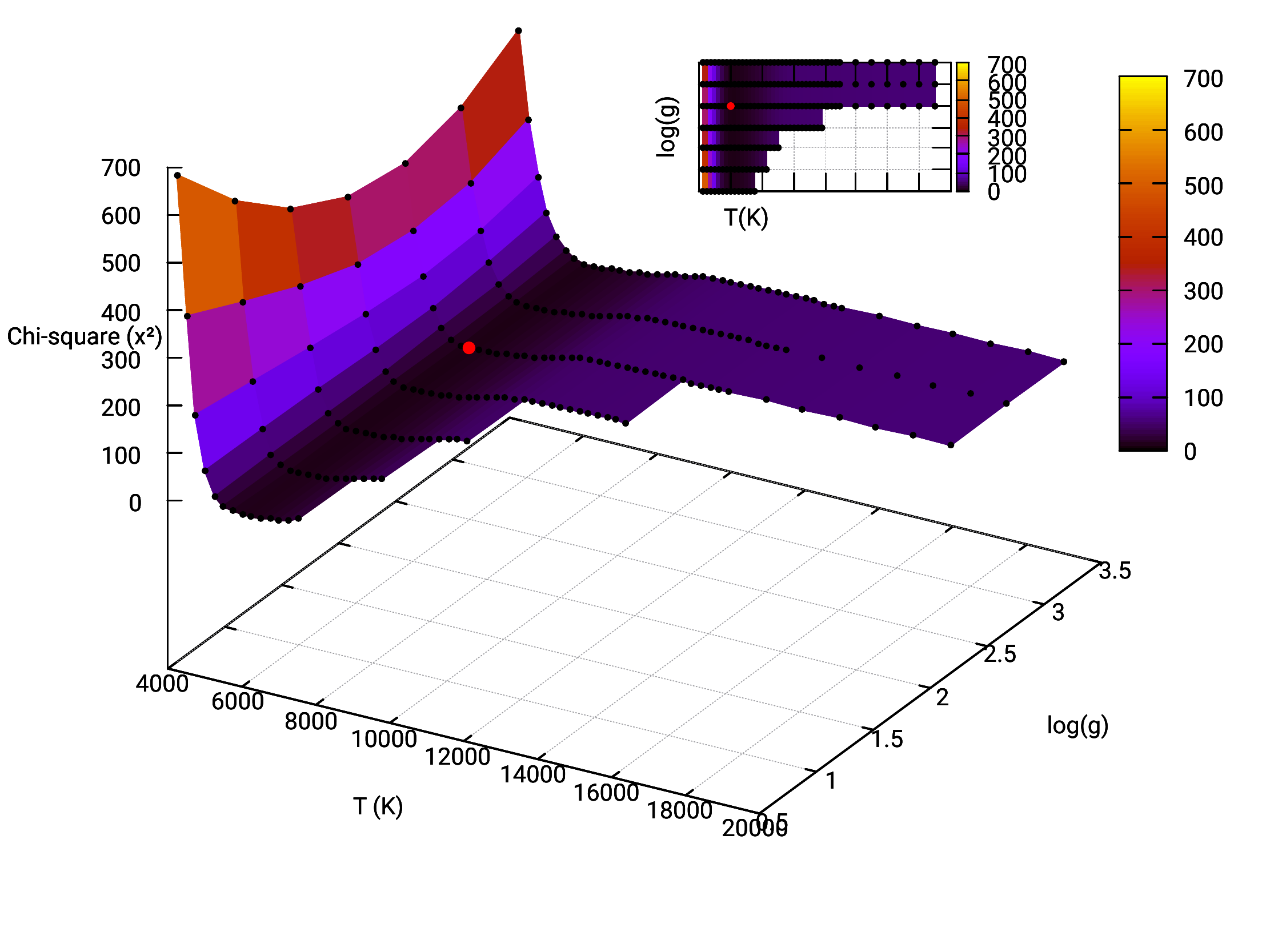}
	\end{center}
	\footnotesize
	{\bf Figure 5.} 
	Colour map of $\chi^2$ with two degrees of freedom and with all other parameters at their optimized values for donor star corresponding to value of $\Delta \chi^2= 8.26$ from a total 12134 data.
	\normalsize 
\end{figure}

Our analysis aimed to find the best synthetic spectrum was based on a chi-square optimization algorithm. The method allowed  simultaneous determination of various parameters involved with stellar spectra and consists of the minimization of the deviation between the theoretical normalized flux distribution and the observed normalized spectrum, the latter corrected by a veiling factor  $\eta$. The synthetic spectrum depends on the stellar parameters, such as effective temperature $T_{eff}$, surface gravity log$g$, rotational velocity $vsini$, micro-turbulence $v_{mic}$ and macro-turbulence velocity $v_{mac}$. The optimization of the method converged successfully at $T_{2}= 6000 \pm 250$ K, log$g_{2}= 2.5 \pm 0.5$, $v_{2r}sini= 26 \pm 1$ km s$^{-1}$, $v_{mic}=0.0 \pm 0.5$ km s$^{-1}$ , $v_{mac}= 11 \pm 0.5 $ (Fig.\,4, 5). We observed a slow divergence of the method for higher temperatures and surface gravity.

%%%%%%%%%%%%%%%%%%%%%%%%%%%%%%%%%%%%%%%%%%%%%%%%%%%%%%%%%%%%%%%%%%%%%%%%%%%%%%%%%%%%%%%%%%%%%%%%%%%%%%%%%%%%%%%%%%%%%%%%%%%%%%%%%%%%%%%
%%%%%%%%%%%%%%%%%%%%%%%%%%%%%%%%%%%%%%%%%%%%%%%%%%%%%%%%%%%%%%%%%%%%%%%%%%%%%%%%%%%%%%%%%%%%%%%%%%%%%%%%%%%%%%%%%%%%%%%%%%%%%%%%%%%%%%%
%%%%%%%%%%%%%%%%%%%%%%%%%%%%%%%%%%%%%%%%%%%%%%%%%%%%%%%%%%%%%%%%%%%%%%%%%%%%%%%%%%%%%%%%%%%%%%%%%%%%%%%%%%%%%%%%%%%%%%%%%%%%%%%%%%%%%%%
\vskip 0.5cm 
{\bf 4.3 Radial velocities for the donor}\\

The radial velocity of the donor was measured by cross correlation of a spectral region plenty  of 
metallic lines with a spectrum chosen as a template. Then the velocities were translated to the absolute 
heliocentric system by adding the template velocity obtained by simple Gaussian fits to selected absorption lines. The donor velocities are given in Table 4.

To find the orbital elements of V495 Cen, we used the genetic algorithm PIKAIA \footnote{http://www.hao.ucar.edu/modeling/pikaia/pikaia.php} developed by \citet{1995ApJS..101..309C}. The method consists in finding the set of orbital parameters that produces a series of theoretical velocities that minimize the function $\chi^{2}$, defined as:\\

$\chi^{2}(P_{o},\tau,\omega,e,K_{2},\gamma)$
\begin{equation}
	= \frac{1}{N-6}\sum_{j=1}^{N} \left(\frac{V_{j}^{obs}-V(t_{j};P_{o},\tau,\omega,e,K_{2},\gamma)}{\sigma_{j}}  \right)^2,\hspace{0.5cm}
\end{equation}
\\
Where \emph{N} is the number of observations, $V_{j}^{obs}$ is the radial velocity observed in the data set, and $V(t_{j}; P_{o}, \tau, \omega, e, K_{2},\gamma)$ is the radial velocity at time $t_{j}$ given the parameters. $P_{o}$ is the orbital period, $\omega$ the periastron longitude, $\tau$ the time of passage per the periastron, \emph{e} the orbital eccentricity, $K_{2}$ the half-amplitude of the RV for the donor, and $\gamma$ the velocity of the system centre of mass. The radial velocity is given by:\\

\begin{equation}
	V(t)=\gamma + K_{2}(cos(\omega+\theta(t)) +e cos(\omega)), \hspace{2.3cm}
\end{equation}
\\

(\citet{2001icbs.book.....H}, eq. 2.45) where $\theta$ is the true anomaly obtained solving the following two equations involving the eccentric anomaly \emph{E}:\\

\begin{equation}
	\tan\left(\frac{\theta}{2}\right)= \sqrt{\frac{1+e}{1-e}}\tan\left(\frac{E}{2}\right), \hspace{3.3cm}
\end{equation}
\\
\begin{equation}
	E-e sin(E) = \frac{2\pi}{P_{o}}(t-\tau), \hspace{4cm}
\end{equation}
\\
(\citet{2001icbs.book.....H}, eq. 2.35) To calculate this, we must first solve equation (6) for \emph{E}, then solve Equation (5) for $\theta$, and finally solve equation (4) for $V(t_{j}; P_{o}, \tau, w, e, K_{2},\gamma)$. The error was estimated using Monte Carlo simulations, by perturbing the best-fitting solution obtained with PIKAIA and computing a $\chi^{2}$ of these perturbed solutions. We note that our eccentricity solutions give a small value of $e=0.007$. We have implemented a test for the significance of the eccentricity (\citet{1971AJ.....76..544L}). According to this test the condition $p_{1} < 0.05$ means a significant ellipticity, where:\\

\begin{equation}
	p_{1}= \left(\frac{\sum (o-c)^{2}_{ecc}}{\sum(o-c)^{2}_{circ}}\right)^{(n-m)/2}
\end{equation}
\\

\noindent
where \emph{ecc} indicates the residuals of an eccentric fit, and \emph{circ} the residuals of a circular fit, \emph{n} the total number of observations, \emph{m} the number of free parameters in an eccentric fit. Finally p$_{1}$ is the probability of falsely rejecting the circular orbit. We obtained $p_{1}=0.0178$  which is in principle compatible with a very small ellipticity. However, the 1$\sigma$ error still allows a circular orbit. The orbital parameters and errors are given in Table 1 and the best fit to the radial velocities is shown in Fig.\,6.

\begin{table*}
\footnotesize
	\caption{Orbital elements for the donor of V495 Cen obtained through minimization of $\chi^{2}$ given  by equation (1). The value $\tau^{*}= \tau-2450000$ is given and the maximum and minimum are one isophote $1\sigma$}
\normalsize
	\[
	\begin{array}{l r r r}
	\hline
	\noalign{\smallskip}
	\textrm {Parameter}   & \textrm{Best value} & \textrm{Low limit} & \textrm{Upper limit}\\
	\hline
	\hline
	P_{o} (d)             &     33.48100  & 33.35620    & 33.60620  \\
	\tau^{*}              & 	105.12600 & 105.02200   & 105.23000 \\
	\emph{e}              & 	0.00733   & 0.00000     & 0.02600   \\
	\omega (rad)          & 	1.68911   & 1.67044     & 1.70844   \\
	K_{2} (km s^{-1})     & 	106.88000 & 105.19200   & 108.49200 \\
	\gamma (km s^{-1})    & 	-0.75273  & -2.12456    & 0.61544  \\
	\hline
	\end{array}
	\]
\end{table*}

The secondary star dominates the metallic and Balmer line spectrum. It was quite difficult to find lines from the gainer. However, we found a little group of characteristic lines that represents the movement of the gainer star and we choose to use the most characteristic lines for a star of B spectral type, these were found as components of the He\,I\,5015 and He\,I\,5875 lines (Fig. 7), as also happened in the DPV HD\,170582 (\citet{2015MNRAS.448.1137M}). It should be noticed that in HD\,170582 the origin of the anomalous He\,I component is placed near the hotspot region, consistent with a hotspot wind.  The radial velocities for the C1 component (gainer) of the  He\,I 5015 line and the C2 component (donor) of the He\,I\,5875 line  were measured with a gaussian fit and are shown in Fig. 8, along with fits for both orbits. We observe that the technique of gaussian fit provides velocities with a larger scatter than those obtained with the cross correlation method.

\begin{figure}
	\begin{center}
		\includegraphics[width=7 cm,angle=0]{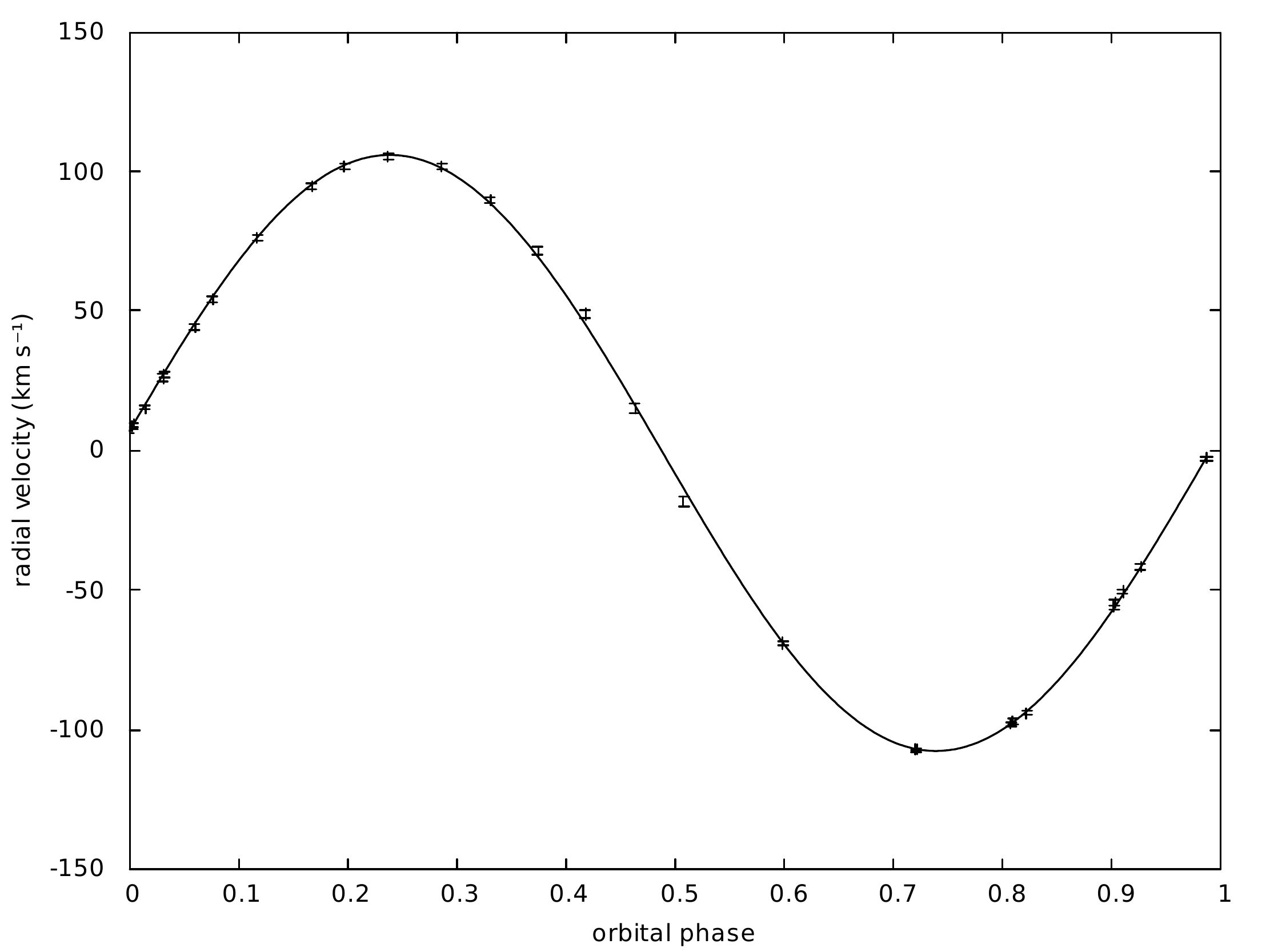}
	\end{center}
	\footnotesize
	{\bf Figure 6} The radial velocities of the donor obtained by cross correlation of a region plenty of metallic lines and the best fit, given for Eq. 4.	
	\normalsize
\end{figure}

\begin{figure}
	\begin{center}
		\includegraphics[width=8 cm,angle=0]{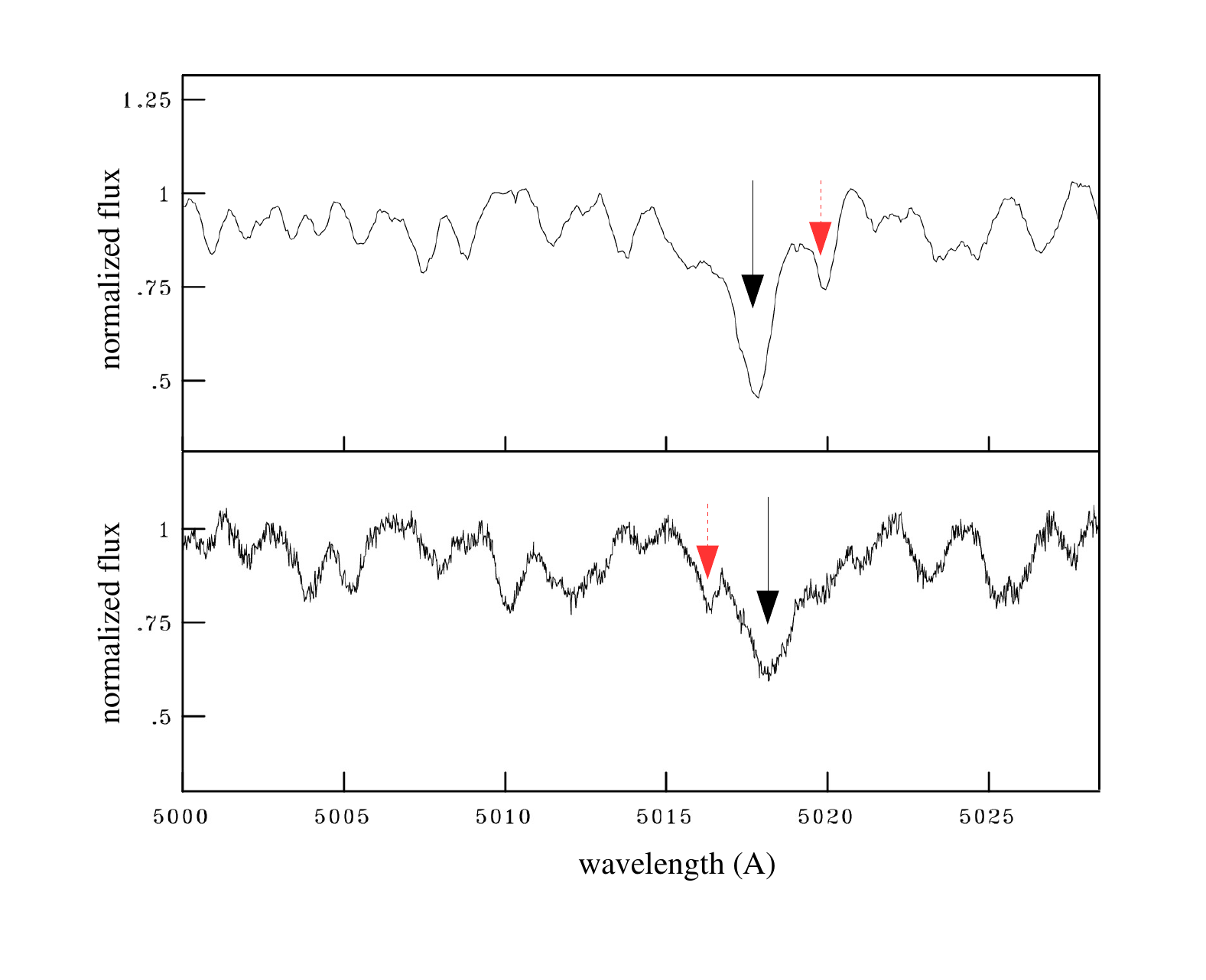}
	\end{center}
	\footnotesize
	{\bf Figure 7.} He I 5015 for HJD 2457062.78265893 ($\Phi_{o}=0.24$, $\Phi_{l}=0.68$, up) and for HJD 2457078.7407534 ($\Phi_{o}=0.72$, $\Phi_{l}=0.70$, below). The dotted (red) arrow indicates the donor star component while the solid (black) arrow indicates the component attributed to the gainer.
	\normalsize
\end{figure}

The RV of the component C1 for He I 5015 is fitted with a sine function of amplitude $16.92 \pm 0.84$ km s$^{-1}$ and zero point $0.00\pm 0.58$ km s$^{-1}$. It moves in opposition to the donor and it could reflect the motion of the gainer around the binary center of mass.  The RV of the component C2 of He\,I\,5875 was fitted with a sine of amplitude $106.80 \pm 0.39$ km s$^{-1}$ and zero point $-0.91\pm 0.28$ km s$^{-1}$, both solutions assume a circular orbit   (Fig. 8). The reason for which the radial velocity curve of the component C1 has a small phase lag might be due to the brightness contribution of the disc in the absorption line.

\begin{figure}
	\begin{center}
\includegraphics[width=8 cm,angle=0]{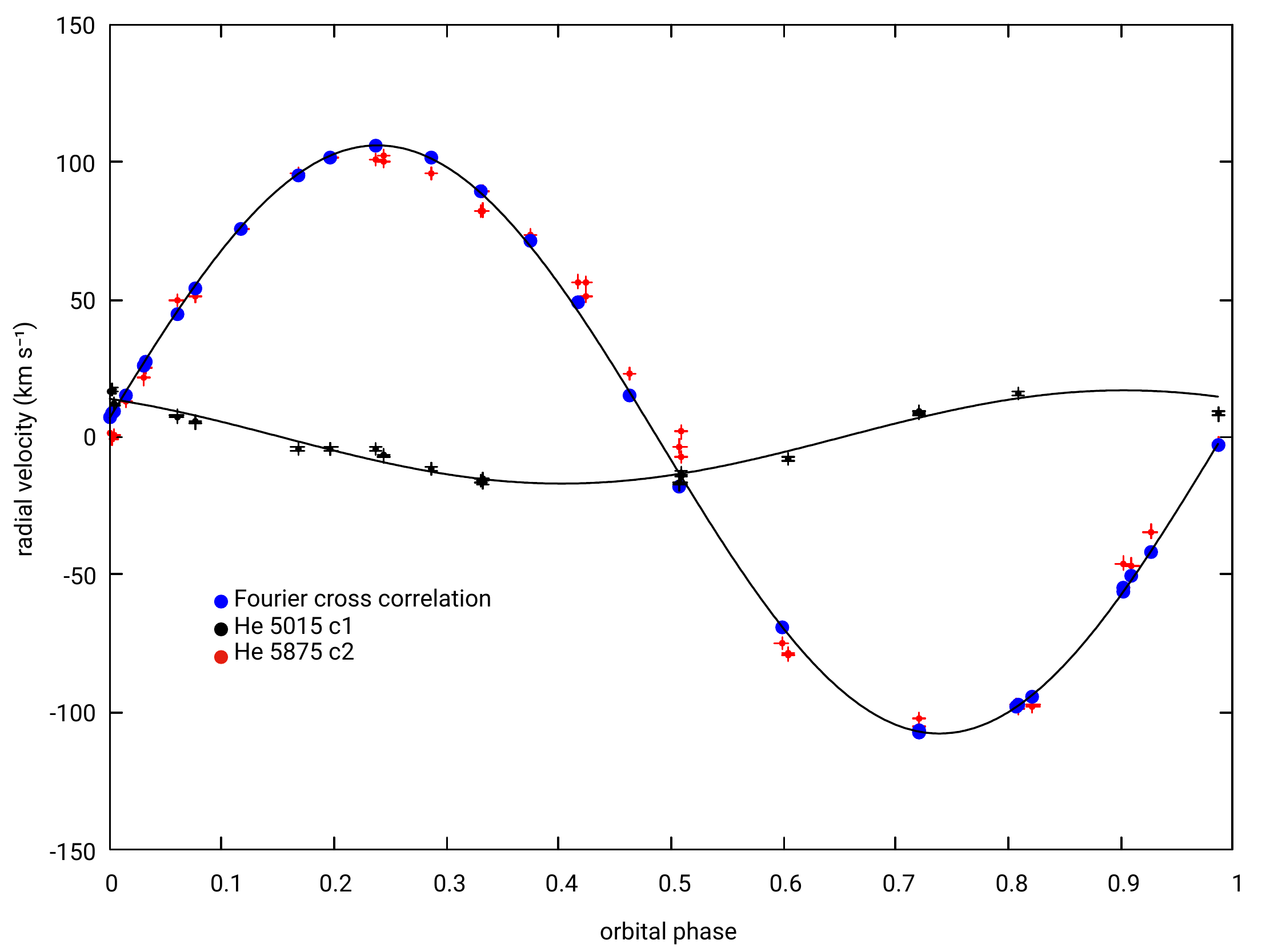}
	\end{center}
\footnotesize
	{\bf Figure 8.} Radial velocities of the He\,I\,5015 C1 component and He\,I\,5875 C2 component measured by gaussian fits. For comparison we show the velocities measured with the  cross correlation method, applied to a region plenty of metallic lines (red dots). The best fits are shown for both orbits. 
	\normalsize
\end{figure}

\vskip 0.5cm
{\bf 4.4 The H$\alpha$ emission line profile}\\

An important result of our spectroscopic analysis is that the H$\alpha$ profile shows a double peak emission with the red peak always of larger intensity than the blue one while  the central absorption does not follow the motion of none of the stellar components (Fig.\,9). Representative H$\alpha$ profiles at some epochs and intensity/width measurements performed at the mean profile are shown in Fig.\,10.
The full width of the line at the level of the continuum is 776 km s$^{-1}$.

The double emission and the deep absorption core is typical of a disk seen at large inclination, however the persistence of the larger red peak is not usual. In binaries with accretion discs, it is normal to find  variations of the relative intensities of the emission peaks. This is quantified with the parameter $V/R$  usually defined as $V/R$ = $(I_v-1)$/$(I_r-1)$, i.e. in terms of the peak intensities relative to the normalized continuum. Due to bright zones in the disc of Algols this ratio tends to vary cyclically with the orbital period.  However, in V495 Cen we observe almost always $V < R$. This can be interpreted as evidence of material escaping from the system, i.e. a wind emerging probably from the hotspot region. We also notice here that the strength of the H$\alpha$ emission is large compared with other cases of DPVs. The equivalent width (EW) of the H$\alpha$ line shows a clear periodic behavior and varies with the orbital period. A least square fit of sine type, with the orbital period fixed at 1.0 P$_{o}$ (solid line in Fig. 11), represents the observed data quite well, and shows a mean EW of 9.517 $\pm$ 0.063 \AA{} and an amplitude of 1.985 $\pm$ 0.663 \AA{}. The phase of observations is such that the minima occur at orbital phase 0.15. A probable interpretation is that there is an asymmetric distribution of H$\alpha$ emission on the surface of the star or material escaping from the hotspot, which is assumed to be synchronously rotating. Also we note the spectral resolution difference for each equivalent width obtained by fiber mode (R$\sim$25000, blue dots) and slicer mode (R$\sim$80000, black dots).

\begin{figure}[htb]
	\begin{center}
		\includegraphics[trim=0.2cm 0.2cm 0.2cm 0.2cm,clip,width=0.41\textwidth,angle=0]{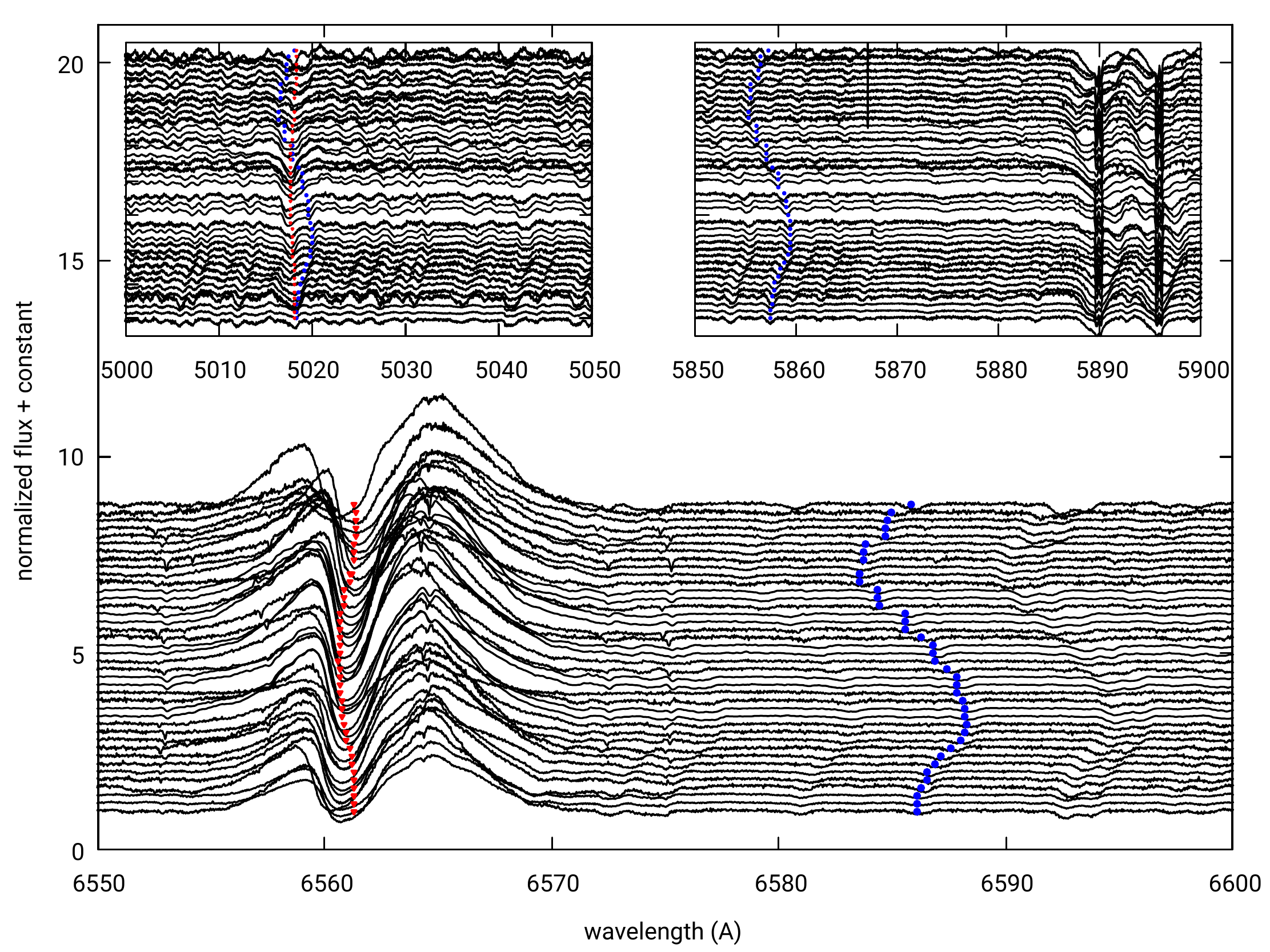}
	\end{center}
	\footnotesize
	{\bf Figure 9.} The behavior of H$\alpha$ line sorted by phases. The radial velocities were over plotted in each spectrum to compare with the movement of the donor (blue dots) and gainer (red triangles).
	\normalsize
\end{figure}

\begin{figure*}[htb]
	\begin{center}
		\includegraphics[width=8 cm,angle=0]{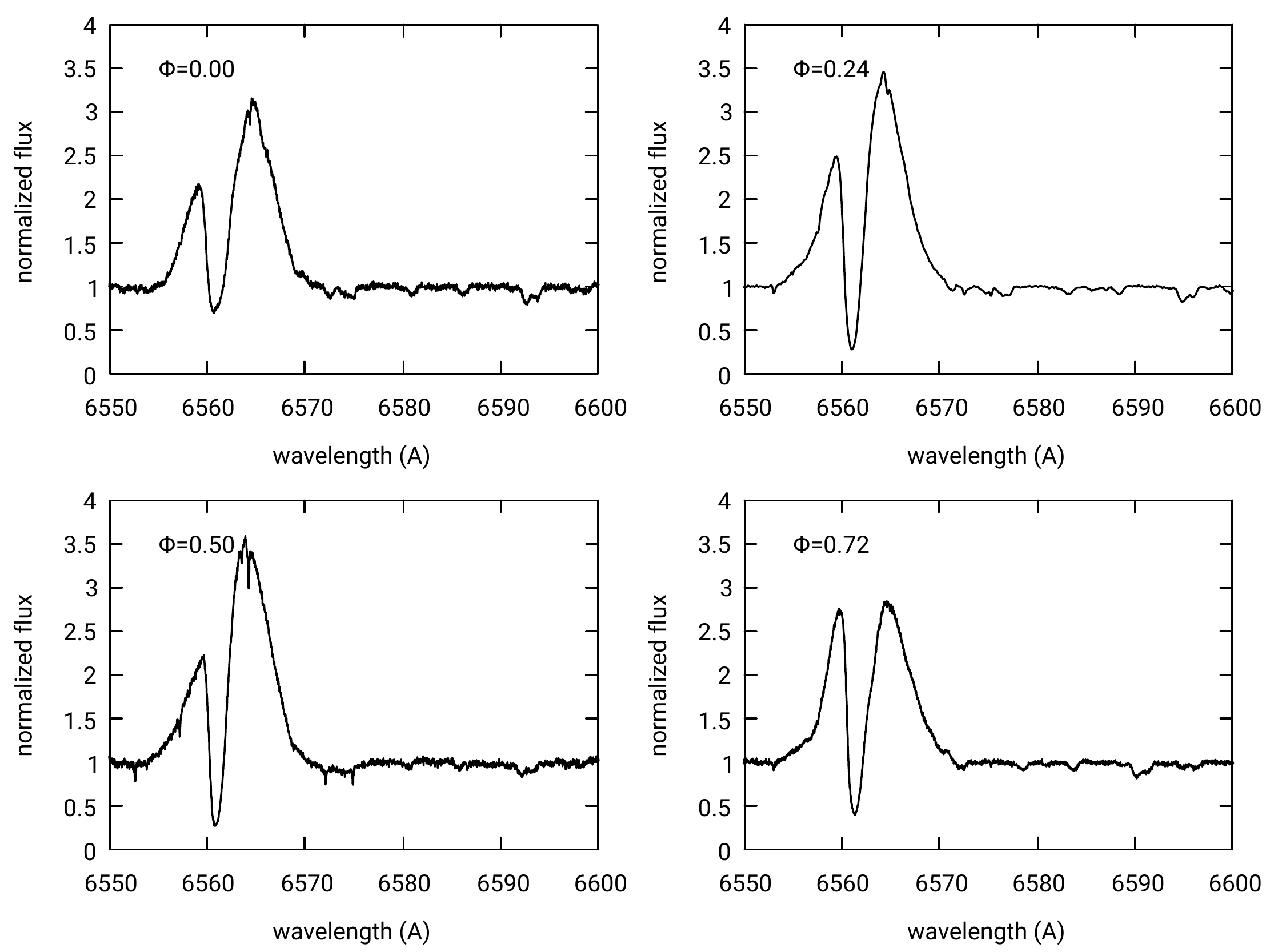}
		\includegraphics[trim=1cm 1cm 1cm 1cm,clip,width=0.47\textwidth,angle=0]{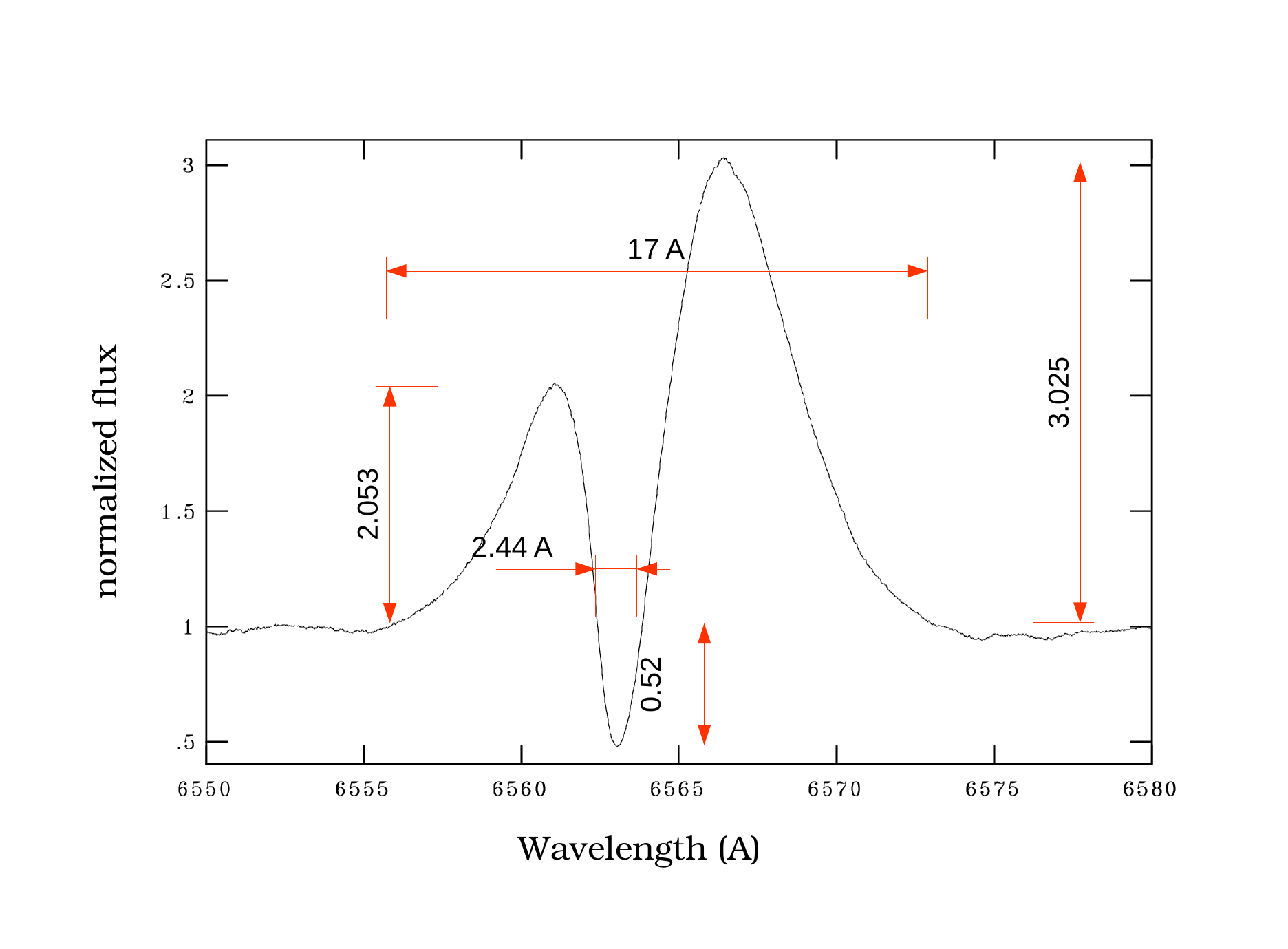}
	\end{center}
	\footnotesize
	{\bf Figure 10.} (Left) Variation of the H$\alpha$ profile at different orbital phases. (Right) H$\alpha$ profile average with representative sizes. The width of the absorption was measured at the height of the continuum but was raised to facilitate the visibility of the data. 
	\normalsize
\end{figure*}

\begin{figure}[htb]
	\begin{center}
		\includegraphics[trim=0.2cm 0.2cm 0.2cm 0.2cm,clip,width=0.41\textwidth,angle=0]{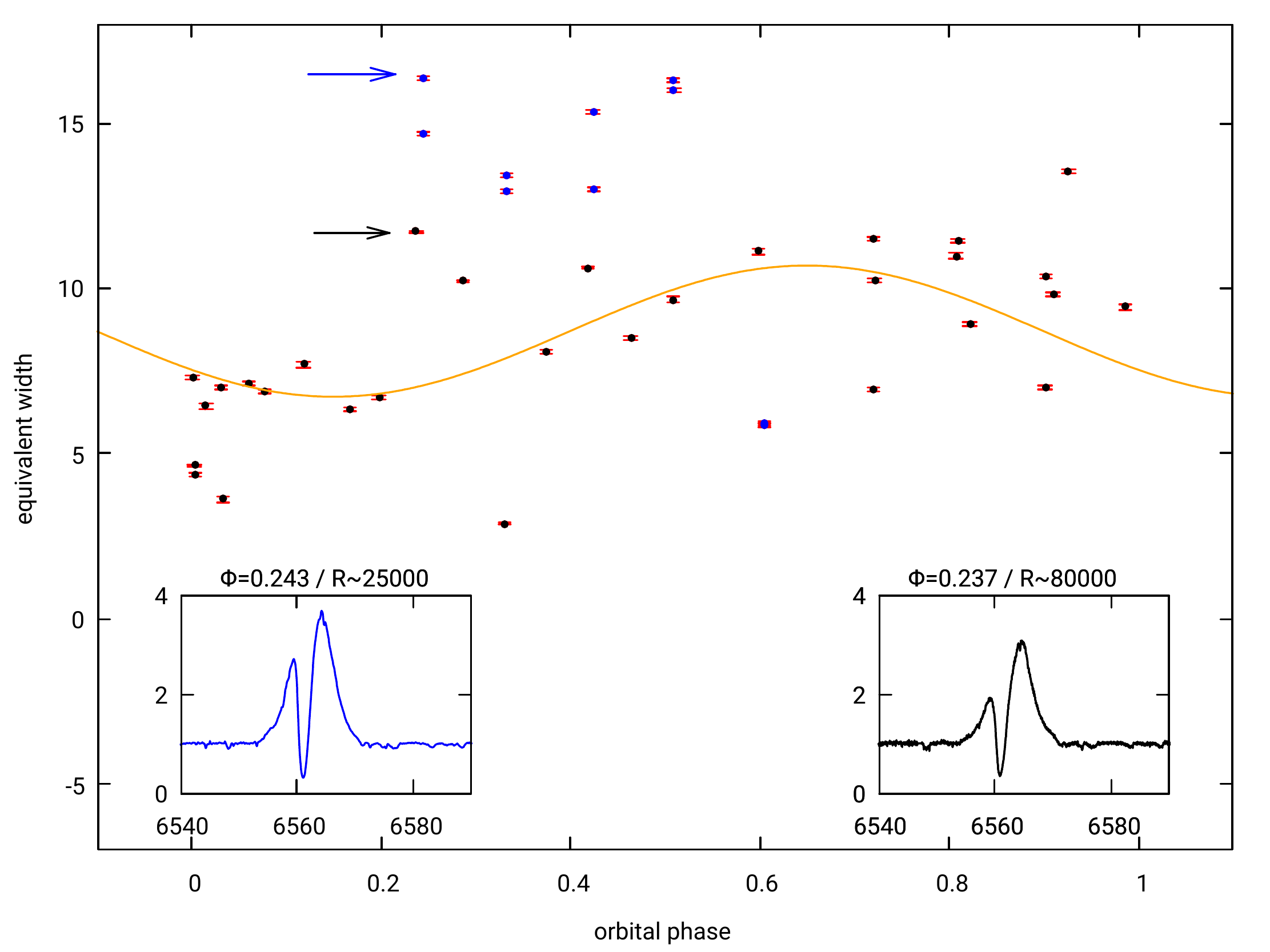}
	\end{center}
\footnotesize
	{\bf Figure 11.} The equivalent width of the H$\alpha$ emission line as a function of the orbital phase. Black dots correspond to spectra obtained with slicer mode, while the blue dots are spectra obtained in fiber mode.
	\normalsize
\end{figure}

%%%%%%%%%%%%%%%%%%%%%%%%%%%%%%%%%%%%%%%%%%%%%%%%%%%%%%%%%%%%%%%%%%%%%%%%%%%%%%%%%%%%%%%%%%%%%%%%%%%%%%%%%%%%%%%%%%%%%%%%%%%%%%%%%%%%%%%
%%%%%%%%%%%%%%%%%%%%%%%%%%%%%%%%%%%%%%%%%%%%%%%%%%%%%%%%%%%%%%%%%%%%%%%%%%%%%%%%%%%%%%%%%%%%%%%%%%%%%%%%%%%%%%%%%%%%%%%%%%%%%%%%%%%%%%%
%%%%%%%%%%%%%%%%%%%%%%%%%%%%%%%%%%%%%%%%%%%%%%%%%%%%%%%%%%%%%%%%%%%%%%%%%%%%%%%%%%%%%%%%%%%%%%%%%%%%%%%%%%%%%%%%%%%%%%%%%%%%%%%%%%%%%%%
\vskip 0.5cm
{\bf 4.5 Gainer, mass ratio and circumstellar matter}\\

In some spectroscopic binaries stars, spectral lines from both stars are visible and the lines are alternately double or single. These systems are known as double-lined spectroscopic binaries, denoted as SB2. But in other systems, the spectrum of only one component is seen and the lines in the spectrum shift periodically towards the blue and red. Such stars are known as single-lined spectroscopic binaries or SB1. V495 Cen is probably an SB2 system and we have  assumed that the velocity obtained from the C1 He\,I 5015 line represents the gainer orbital motion hence we have inferred a mass ratio $q=0.158 \pm 0.008$.

Now let's investigate if this value is compatible with synchronous rotation for the secondary star filling the Roche lobe; we use the following equation valid for semi detached binaries:\\

\begin{equation}
\frac{v_{rot}sini}{K} \approx (1+q) \frac{0.49 q^{2/3}}{0.6q^{2/3}+ln(1+q^{1/3})}
\end{equation}
\\

\noindent
(\citet{2006epbm.book.....E}, eq. 3.5 and 3.9). Therefore with $K_{2}=106.880$ km {s${}^{-1}$} and $v_{rot2}sin i= 26$ km s${}^{-1}$ with their respective errors and considering the $1\%$ of accuracy of equation 8, we get $q$= 0.119 $\pm$ 0.037, which is very close to the $q$ value derived from the RV half-amplitude of the helium components (Fig.\,12); therefore we can say that the 5015 HeI line comes from the gainer star. 
Since synchronism is expected for such a short orbital period binary due to the influence of dynamical tides (\citet{1975A&A....41..329Z,1977A&A....57..383Z}),  from now on we assume the donor rotating synchronously. 

Another important point is the relation between the mean density $\overline{\rho}$ (in solar units) of a star that just fills its Roche lobe and the effective radius R$_{L}$ (\citet{1983ApJ...268..368E}) we quantified as the product between the critical orbital period (days) and mean density (g cm$^{-3}$) as:\\

\begin{equation}
	P_{cr} \sqrt{\overline{\rho}}= \left(\frac{3\pi}{G}\right)^{1/2}\left(\frac{q}{1+q}\right)^{1/2} x_{L}^{-3/2} 
\end{equation}
\\

\noindent
where $x_{L}$ is in units of the orbital separation. Assuming $q=0.158$ from the above RV study, we have obtained the ratio between the effective radius and the orbital separation $R_{L}/a=0.236 \pm 0.155$ (\citet{2006epbm.book.....E}, eq. 3.5), and estimated the critical period $P_{cr}\sqrt{\overline{\rho}}= 0.44 \pm 0.06$, which is the shortest period possible for a binary of given mass ratio into which a star of given  mean density $\overline{\rho}$ can be fitted without overflowing its Roche lobe (\citet{1983ApJ...268..368E}, eq. 3). Both values are significant figures to understand the evolution of the Double Periodic Variable V495 Cen.

\begin{figure}
	\begin{center}
		\includegraphics[width=8 cm,angle=0]{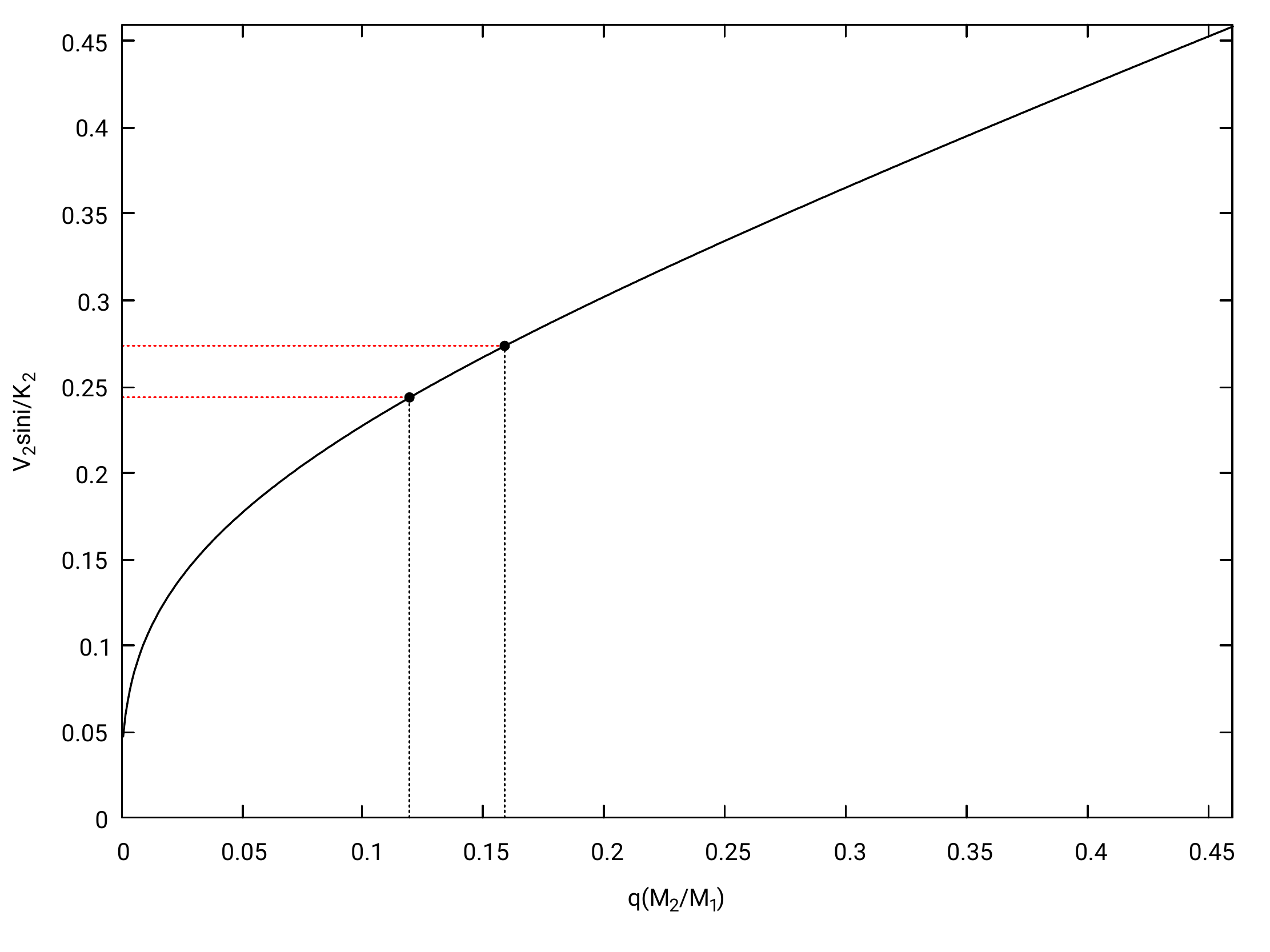}
	\end{center}
	\footnotesize
	{\bf Figure 12.} The solid line is given by equation 8 and the dashed lines show the synchronous ($q=0.158$) and the observed (sub-synchronous, $q=0.119$) cases.
	\normalsize
\end{figure}
%%%%%%%%%%%%%%%%%%%%%%%%%%%%%%%%%%%%%%%%%%%%%%%%%%%%%%%%%%%%%%%%%%%%%%%%%%%%%%%%%%%%%%%%%%%%%%%%%%%%%%%%%%%%%%%%%%%%%%%%%%%%%%%%%%%%%%%
%%%%%%%%%%%%%%%%%%%%%%%%%%%%%%%%%%%%%%%%%%%%%%%%%%%%%%%%%%%%%%%%%%%%%%%%%%%%%%%%%%%%%%%%%%%%%%%%%%%%%%%%%%%%%%%%%%%%%%%%%%%%%%%%%%%%%%%
%%%%%%%%%%%%%%%%%%%%%%%%%%%%%%%%%%%%%%%%%%%%%%%%%%%%%%%%%%%%%%%%%%%%%%%%%%%%%%%%%%%%%%%%%%%%%%%%%%%%%%%%%%%%%%%%%%%%%%%%%%%%%%%%%%%%%%%
\vskip 0.5cm
{\bf 4.6 Mass constraints from spectroscopy}\\

The mass function represents the minimum possible mass for the unseen star, for a system with unknown orbital inclination $i$ and elliptical orbit, it may be expressed as:\\

\footnotesize
\begin{equation}
f= 1.0361 \times 10^{-7} (1-e^{2})^{3/2} \left( \frac{K_{2}}{\rm{km s^{-1}}}\right)^{3} \frac{P_{o}}{\rm{day}} \rm{M_{\astrosun}}
\end{equation}
\normalsize
\\

\noindent
(Hilditch 2001, eq. 2.53)
Using $q = 0.158$ (donor rotating synchronously) and an elliptical orbit with $e$= 0.007 from the above RV study, we obtained a mass function $f=4.235 \pm 0.015 M_{\astrosun}$. Assuming an orbital inclination $i = 84.8$ (see the next section), the respective masses are $M_{d}= 0.911 \pm 0.420 M_{\astrosun}$ and $M_{g}= 5.753 \pm 0.451 M_{\astrosun}$. These are preliminary results. Both are compatible with results obtained, independently, through LC analysis in the next section.\\
 
%%%%%%%%%%%%%%%%%%%%%%%%%%%%%%%%%%%%%%%%%%%%%%%%%%%%%%%%%%%%%%%%%%%%%%%%%%%%%%%%%%%%%%%%%%%%%%%%%%%%%%%%%%%%%%%%%%%%%%%%%%%%%%%%%%%%%%%
%%%%%%%%%%%%%%%%%%%%%%%%%%%%%%%%%%%%%%%%%%%%%%%%%%%%%%%%%%%%%%%%%%%%%%%%%%%%%%%%%%%%%%%%%%%%%%%%%%%%%%%%%%%%%%%%%%%%%%%%%%%%%%%%%%%%%%%
%%%%%%%%%%%%%%%%%%%%%%%%%%%%%%%%%%%%%%%%%%%%%%%%%%%%%%%%%%%%%%%%%%%%%%%%%%%%%%%%%%%%%%%%%%%%%%%%%%%%%%%%%%%%%%%%%%%%%%%%%%%%%%%%%%%%%%%
\vskip 0.5cm 
{\bf 5 LIGHT-CURVE MODEL AND SYSTEM PARAMETERS}\\

\vskip 0.5cm
{\bf 5.1 The fitting procedure}\\

We find the main physical parameters for the stellar components fitting the V-band light curve with the aid of an inverse-problem solving method based in the algorithm developed by \citet{1992Ap&SS.197...17D,1996Ap&SS.243..413D}. We have assumed a semi-detached system with the donor filling the Roche lobe, and we adopted a configuration that included an optically thick accretion disc around the gainer star.

Our model considers a hot spot located on the edge side of the disc, and it is described by the ratio of the hot spot temperature and the unperturbed local temperature of the disc, the angular dimension and longitude (in arc degrees). Synchronous rotation is assumed for both stellar components. The model used follows results of hydrodynamical simulation of gas dynamics in interacting close binary stars by \citet{1998MNRAS.300...39B,1999ARep...43..797B,2003ARep...47..809B} and has been tested in studies of DPVs (e.g. \citet{2012MNRAS.421..862M},\citet{2013A&A...552A..63B},\citet{2015IAUS..307..125M}). 

We fixed $q=0.158$ and $T_{c}=6000 K$ based on the spectroscopic study previously presented. In addition, we set the gravity darkening exponent and the albedo of the gainer to $\beta_{h}= 0.25$ and $A_{h}=1.0$ in accordance with von Zeipel's law for radiative shells and complete re-radiation; for the donor we set $\beta_c=0.08$ and $A_c=0.5$, as is appropriate for stars with convective envelopes according to \citet{1967ZA.....65...89L},\citet{1980MNRAS.193...79R},\citet{1969AcA....19..245R}.

\begin{table}
\footnotesize

\caption{Results of the analysis of the $V$-band light-curve of V495\,Cen obtained by solving the inverse problem for the Roche model with an accretion disc around the  more-massive (hotter) gainer in the synchronous rotation regime.}
\normalsize
\label{TabV393Sco}
\[
\begin{array}{llll}
\hline
\noalign{\smallskip}
{\textrm {Quantity}} & & {\textrm {Quantity}} & \\
\noalign{\smallskip}
\hline
\noalign{\smallskip}
n                                  & 617             		& \cal M_{\rm_h} {[\cal M_{\odot}]} & 5.76  \pm 0.3  \\
{\rm \Sigma(O-C)^2}                & 0.5751          		& \cal M_{\rm_c} {[\cal M_{\odot}]} & 0.91  \pm 0.2  \\
{\rm \sigma_{rms}}                 & 0.0306          		& \cal R_{\rm_h} {\rm [R_{\odot}]}  & 4.5   \pm 0.2  \\
i {\rm [^{\circ}]}                 & 84.8  \pm 0.6   	& \cal R_{\rm_c} {\rm [R_{\odot}]}  & 19.3  \pm 0.5  \\
{\rm F_d}                          & 0.88  \pm 0.03  	& {\rm log} \ g_{\rm_h}             & 3.89  \pm 0.02  \\
{\rm T_{d}} [{\rm K}]                & 4040  \pm 250   	& {\rm log} \ g_{\rm_c}             & 1.83  \pm 0.02  \\
{\rm d_e} [a_{\rm orb}]            & 0.046 \pm 0.016 	& M^{\rm h}_{\rm bol}               &-3.16  \pm 0.1  \\
{\rm d_c} [a_{\rm orb}]            & 0.007 \pm 0.009 	& M^{\rm c}_{\rm bol}               &-1.80  \pm 0.1  \\
{\rm a_T}                          & 4.1   \pm 0.3   	& a_{\rm orb}  {\rm [R_{\odot}]}    & 82.8  \pm 0.3  \\
{\rm f_h}                          & 1.00            		& \cal{R}_{\rm d} {\rm [R_{\odot}]} & 40.2  \pm 1.3  \\
{\rm F_h}                          & 0.109 \pm 0.014 	& \rm{d_e}  {\rm [R_{\odot}]}       & 3.8   \pm 0.2  \\
{\rm T_h} [{\rm K}]                & 16960 \pm 400   	& \rm{d_c}  {\rm [R_{\odot}]}       & 0.6   \pm 0.2  \\
{\rm A_{hs}=T_{hs}/T_{d}}         & 1.11  \pm 0.05  	&                                                   \\
{\rm \theta_{hs}}{\rm [^{\circ}]}  & 18.2  \pm 2.0   		&                                                   \\
{\rm \lambda_{hs}}{\rm [^{\circ}]} & 338.3 \pm 9.0   		&                                                   \\
{\rm \theta_{rad}}{\rm [^{\circ}]} & -15.4  \pm 13.6 	&                                                   \\
{\Omega_{\rm h}}                   & 18.36  \pm 0.02	 	&                                                   \\
{\Omega_{\rm c}}                   & 2.125  \pm 0.05 	&                                                   \\
\noalign{\smallskip}
\hline
\end{array}
\]
\footnotesize
\emph{Note}: \emph{Fixed Parameters}: $q={\cal M}_{\rm c}/{\cal M}_{\rm h}=0.158$ - mass ratio of the components, ${\rm T_c=6000 K}$ -temperature of the less-massive (cooler) donor, ${\rm F_c}=1.0$ -filling factor for the critical Roche lobe of the donor, $f{\rm _{h,c}}=1.00$ - non-synchronous rotation coefficients of the system components, ${\rm \beta_h=0.25}$, ${\rm \beta_c=0.08}$- gravity-darkening coefficients of the components, ${\rm A_h=1.0}$,${\rm A_c=0.5}$ - albedo coefficients of the components.

\emph{Note}: $n$ - number of observations, ${\rm\Sigma (O-C)^2}$ - final sum of squares of residuals between observed (LCO) and synthetic (LCC) light-curves, ${\rm\sigma_{rms}}$ - root-mean-square of the residuals, $i$ - orbit inclination (in arc degrees), ${\rm F_d=R_d/R_{yc}}$ - disk dimension factor (the ratio of the disk radius to the critical Roche lobe radius along y-axis), ${\rm T_{d}}$ - disk-edge temperature, $\rm{d_e}$, $\rm{d_c}$,  - disk thicknesses (at the edge and at the center of the disk, respectively) in the units of the distance between the components, $a_{\rm T}$ - disk temperature distribution coefficient, $f{\rm _h}$ - non-synchronous rotation coefficient of the more massive gainer (in the synchronous rotation regime), ${\rm F_h}=R_h/R_{zc}$ - filling factor for the critical Roche lobe of the hotter, more-massive gainer (ratio of the stellar polar radius to the critical Roche lobe radius along z-axis for a star in synchronous rotation regime), ${\rm T_h}$ - temperature of the gainer, ${\rm A_{hs,bs}=T_{hs,bs}/T_{d}}$ - hot spot temperature coefficients, ${\rm \theta_{hs}}$ and ${\rm \lambda_{hs}}$ - spot angular dimension and longitude (in arc degrees), ${\rm \theta_{rad}}$ - angle between the line perpendicular to the local disk edge surface and the direction of the hot-spot maximum radiation, ${\Omega_{\rm h,c}}$ - dimensionless surface potentials of the hotter gainer and cooler donor, $\cal M_{\rm_{h,c}} {[\cal M_{\odot}]}$, $\cal R_{\rm_{h,c}} {\rm [R_{\odot}]}$ - stellar masses and mean radii of stars in solar units, ${\rm log} \ g_{\rm_{h,c}}$ - logarithm (base 10) of the system components effective gravity, $M^{\rm {h,c}}_{\rm bol}$ - absolute stellar bolometric magnitudes, $a_{\rm orb}$ ${\rm [R_{\odot}]}$, $\cal{R}_{\rm d} {\rm [R_{\odot}]}$, $\rm{d_e} {\rm [R_{\odot}]}$, $\rm{d_c} {\rm [R_{\odot}]}$ - orbital semi-major axis, disk radius and disk thicknesses at its edge and center, respectively, given in solar units.\\
\normalsize
\end{table}

%%%%%%%%%%%%%%%%%%%%%%%%%%%%%%%%%%%%%%%%%%%%%%%%%%%%%%%%%%%%%%%%%%%%%%%%%%%%%%%%%%%%%%%%%%%%%%%%%%%%%%%%%%%%%%%%%%%%%%%%%%%%%%%%%%%%%%%
%%%%%%%%%%%%%%%%%%%%%%%%%%%%%%%%%%%%%%%%%%%%%%%%%%%%%%%%%%%%%%%%%%%%%%%%%%%%%%%%%%%%%%%%%%%%%%%%%%%%%%%%%%%%%%%%%%%%%%%%%%%%%%%%%%%%%%%
%%%%%%%%%%%%%%%%%%%%%%%%%%%%%%%%%%%%%%%%%%%%%%%%%%%%%%%%%%%%%%%%%%%%%%%%%%%%%%%%%%%%%%%%%%%%%%%%%%%%%%%%%%%%%%%%%%%%%%%%%%%%%%%%%%%%%%%

\vskip 0.5cm	
{\bf 5.2 The best light-curve model}\\

The best fit model for V495 Cen contains an optically and geometrically thick disc around the gainer star; the stellar, orbital and disc parameters are given in Table 2.

The system inclination angle is  $i=84^{\circ}.8 \pm 0.6$ and the disc radius  $R_{d}=40.2 \pm 1.3 R_{\astrosun}$, which is 9 times larger than the radius of the main star  $R_{h}=4.5$. The disc has central vertical thickness $d_{c}=0.6 R_{\astrosun}$ and edge thickness $d_{e}=3.8 R_{\astrosun}$, i.e. it has a concave form. The temperature of the disc increases from $T_{d}=4040$ K at the outer edge to $T_{h}= 16960$ K at the inner edge where it is in thermal and physical contact with the gainer. The surface gravity of the secondary component (donor) is log$g_{c}=1.83\pm 0.1$  i.e. almost $1\sigma$ below the spectroscopic value $2.5 \pm 0.5$. The temperature of the hotspot is $T_{hs}\approx 4484$ K i.e. is 11 percent higher than the disc edge temperature.

The light curve and the fit, the O-C curve, residuals, and the individual flux contributions of the donor, disc and the gainer, are shown in Fig.\,13. Also we show a view of the model at orbital phases 0.05, 0.55 and 0.80. 

The obtained solution can be treated as unique with a model of the system with  fixed initial parameters (mass ratio and temperature of the donor). The uniqueness of the obtained optimal solution is checked by the solving the inverse problem of the light curve interpretation by applying the Simplex algorithm. Even, if we vary the initial system parameters in the initial simplex in the interval of 5-10$\%$ of the optimal solution, the optimization process converges to the obtained solution within a given errors bars. Of course, if the initial fixed parameters are used with a larger uncertainty, the estimated errors bars of the free model parameters increase too.

\begin{figure}
	\begin{center}
		\includegraphics[width=8 cm,angle=0]{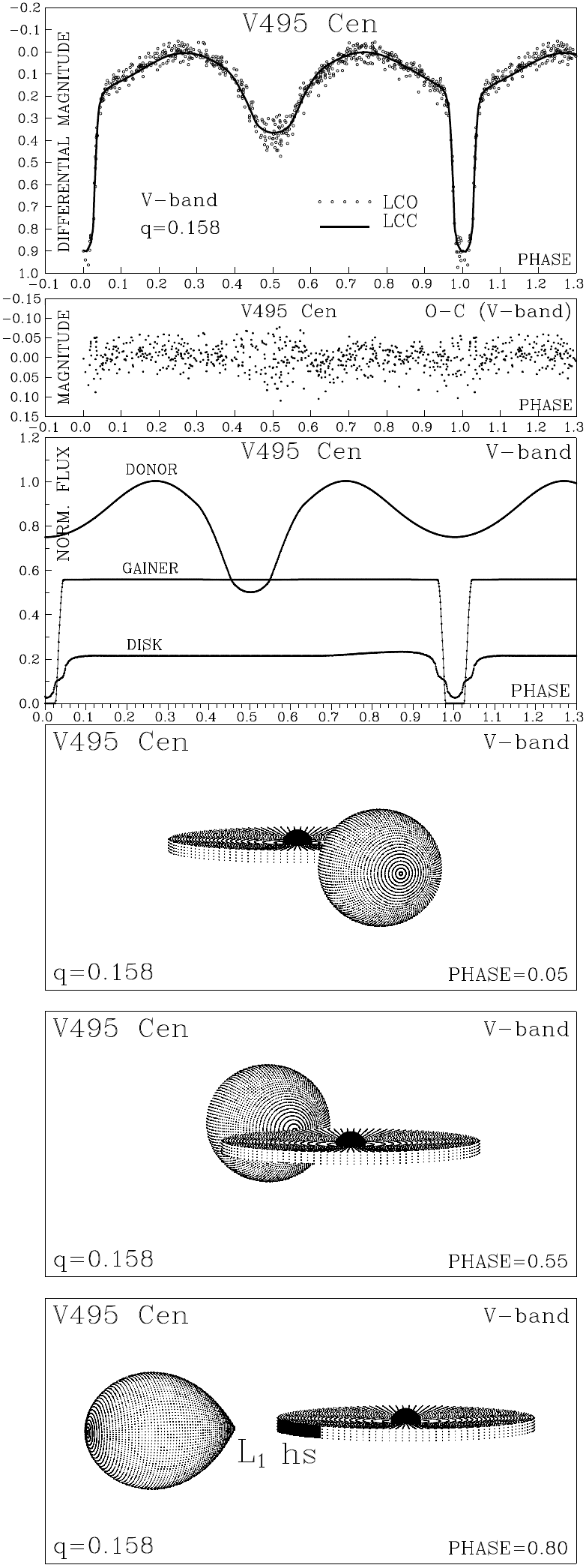}
	\end{center}
\footnotesize
	{\bf Figure 13.} From top to bottom we show the observed (LCO) and synthetic (LCC) light-curves of V495\,Cen obtained by analyzing photometric observations, final O-C residuals between the observed and optimum synthetic light curves, flux of donor, gainer and of the accretion disc, normalized to the donor flux at phase 0.25. The views of the optimal model at orbital phases 0.05, 0.50 and 0.80 obtained with parameters estimated by the light curve analysis.
\normalsize 
\end{figure}

%%%%%%%%%%%%%%%%%%%%%%%%%%%%%%%%%%%%%%%%%%%%%%%%%%%%%%%%%%%%%%%%%%%%%%%%%%%%%%%%%%%%%%%%%%%%%%%%%%%%%%%%%%%%%%%%%%%%%%%%%%%%%%%%%%%%%%%
%%%%%%%%%%%%%%%%%%%%%%%%%%%%%%%%%%%%%%%%%%%%%%%%%%%%%%%%%%%%%%%%%%%%%%%%%%%%%%%%%%%%%%%%%%%%%%%%%%%%%%%%%%%%%%%%%%%%%%%%%%%%%%%%%%%%%%%
%%%%%%%%%%%%%%%%%%%%%%%%%%%%%%%%%%%%%%%%%%%%%%%%%%%%%%%%%%%%%%%%%%%%%%%%%%%%%%%%%%%%%%%%%%%%%%%%%%%%%%%%%%%%%%%%%%%%%%%%%%%%%%%%%%%%%%%
\vskip 0.5cm 
{\bf 6 REDDENING, DISTANCE  AND  SPECTRAL ENERGY DISTRIBUTION}\\

\vskip 0.5cm
{\bf 6.1 Distance determination}\\

In order to determine the distance to the system we have applied a standard method based on the distance modulus  to both binary components observed in the $V$ band (Clausen, 2004):

\begin{eqnarray}
	(m_{d,g}-M_{d,g})_{0} & = &  5log(R_{d,g}/R_{\astrosun})+(m_{d,g}-A_{V})\nonumber\\
	& &- M_{bol\astrosun}+ 10log(T_{d,g}/T_{\astrosun})\nonumber\\
	& & + BC_{d,g}
\end{eqnarray}
\\
where $A_{V}$ is interstellar absorption and $BC$ the bolometric correction; the apparent magnitudes, radii and effective temperatures of gainer and donor are represented with sub-indexes g and d, respectively. 

Maps of Galactic dust and extinction in the region of V495 Cen are available on-line and give  \textrm{$E(B-V)_{S\&F}= 0.2820\pm 0.0069$} (\citet{2011ApJ...737..103S}) and \textrm{$E(B-V)_{SFD}= 0.3279\pm0.0080$} (\citet{1998ApJ...500..525S}).  Assuming a visual extinction to reddening ratio \textrm{$A_{V}/E(B-V)= 3.1$} we get  \textrm{$A_{V_{S\&F}}=0.8460  \pm 0.006$} and \textrm{$A_{V_{SFD}}=0.9837  \pm 0.008$}, where we preferred the most recent determination of the reddening in the direction of the target.

The model of the $V$-band light curve (Fig.\,13) shows that the flux of the system at the $V$ band at quadratures $\phi_{o}=0.25$ and $\phi_{o}=0.75$ is $f_{t}=1.77 \pm 0.05$ while the individual flux contribution of the donor is $f_{d}=1.0 \pm 0.05$, of the gainer is $f_{g}=0.57\pm 0.05$ and the disc is $f_{disc}=0.2\pm 0.05$. This means that at quadrature the donor, gainer and disc contribute 
57, 32 and 11 \% to the total flux at the $V$ band, respectively.

The apparent magnitude of the donor and gainer are derived from:

\begin{equation}
	m_{d}-m_{t}= -2.5 log\left(\frac{f_{d}}{f_{t}}\right)
\end{equation}
\\

Considering that the observed apparent magnitude of the system is $m_{t}(V)= 9.95 \pm 0.03$, and the flux fractions given above, we find m$_{d}$=10.569 $\pm$ 0.044 mag and m$_{g}$=11.180 $\pm$ 0.037 mag. 

The bolometric corrections were taken from \citet{1996ApJ...469..355F} and the solar bolometric magnitude was taken from \citet{2010AJ....140.1158T}. We use logT$_{d}$= 3.778 $\pm$ 0.042 and  BC$_{d}$= -0.045 $\pm$ 0.110 for the donor and logT$_{g}$ =4.229 $\pm$ 0.024 and BC$_{g}$= -1.525 $\pm$ 0.079 for the gainer, and the individual distance are $d_{d}=1923.092 \pm 137.538 \,pc$ $d_{g}= 2401.043 \pm 187.169 \,pc$. The difference in the distance for donor and gainer can be interpreted as an estimate of the intrinsic error of the method due to the gainer star is hidden by the disc, causing an extra reddening that is not the same reddening caused by the total absorption across the line of sight to V495 Cen. Therefore we have averaged the values obtaining for the system distance:

\begin{eqnarray}
	%d(pc)&= & 10^{((m-M)_{o}+5)/5}\nonumber\\
	d(pc)&= & 2162.068\pm 324.707 \, pc
\end{eqnarray}

Since the maximum absorption through the galactic gas column along the system line of sight has been considered for this calculation, this distance must be considered as a lower limit only.

%%%%%%%%%%%%%%%%%%%%%%%%%%%%%%%%%%%%%%%%%%%%%%%%%%%%%%%%%%%%%%%%%%%%%%%%%%%%%%%%%%%%%%%%%%%%%%%%%%%%%%%%%%%%%%%%%%%%%%%%%%%%%%%%%%%%%%%%
%%%%%%%%%%%%%%%%%%%%%%%%%%%%%%%%%%%%%%%%%%%%%%%%%%%%%%%%%%%%%%%%%%%%%%%%%%%%%%%%%%%%%%%%%%%%%%%%%%%%%%%%%%%%%%%%%%%%%%%%%%%%%%%%%%%%%%%%
%%%%%%%%%%%%%%%%%%%%%%%%%%%%%%%%%%%%%%%%%%%%%%%%%%%%%%%%%%%%%%%%%%%%%%%%%%%%%%%%%%%%%%%%%%%%%%%%%%%%%%%%%%%%%%%%%%%%%%%%%%%%%%%%%%%%%%%%
\vskip 0.5cm
{\bf 6.2 Position in the H-R Diagram}\\

From the above sections it is clear that the less massive star is the more evolved star of the binary pair, as usual in Algol systems. This can be explained by the mass transfer that has inverted the system mass ratio in such a way that the less massive star started as the more massive one some time ago and evolved first
until filling its Roche lobe just before start mass transfer onto the now more massive stellar component.
  
In order to study the position of V\,495\,Cen in the luminosity-temperature diagram, we have chosen as comparison models based on a grid that is part of the large database of Geneva stellar  models\footnote{https://obswww.unige.ch/Recherche/evoldb/index/}. We consider single star models with metallicity $Z$= 0.014 (\citet{2012A&A...537A.146E}) as a first approximation to understand the evolutionary stage of V\,495\,Cen (Fig. 14).

We observe that the gainer (5.76 $ M_{\odot}$) is located outside the main sequence, showing a luminosity similar to a single star of that mass. It is located near the start of the stellar contraction stage for the evolutionary track of 6.0 $M_{\odot}$. The core fractions of hydrogen and helium are 0.72 and 0.266, respectively. On the other hand, the donor (0.91 $M_{\odot}$) has a huge luminosity compared with a star of  similar mass and fits the evolutionary track of a 4.35 $M_{\odot}$ single star in the phase of hydrogen shell narrowing (Iben 1967, Fig.14). The high luminosity donor can be understood in terms of an evolved giant filling its Roche lobe that has transferred part of its mass onto the gainer forming the accretion disc.

Also we have inferred the color of each star as function of effective temperature (\citet{2010AJ....140.1158T}). ${(B-V)_{o}}_g=-0.200 \pm 0.001$, ${(B-V)_{o}}_{d}=0.571 \pm 0.001$ and compared the intrinsic color of both obtaining the spectral classes of each star. Hence, considering all above we find that the gainer is an early B3\,V type, and the donor corresponds to late F9\,IV, based on the intrinsic color analysis (\citet{1970A&A.....4..234F}).

\begin{figure}
	\begin{center}
		\includegraphics[width=8 cm,angle=0]{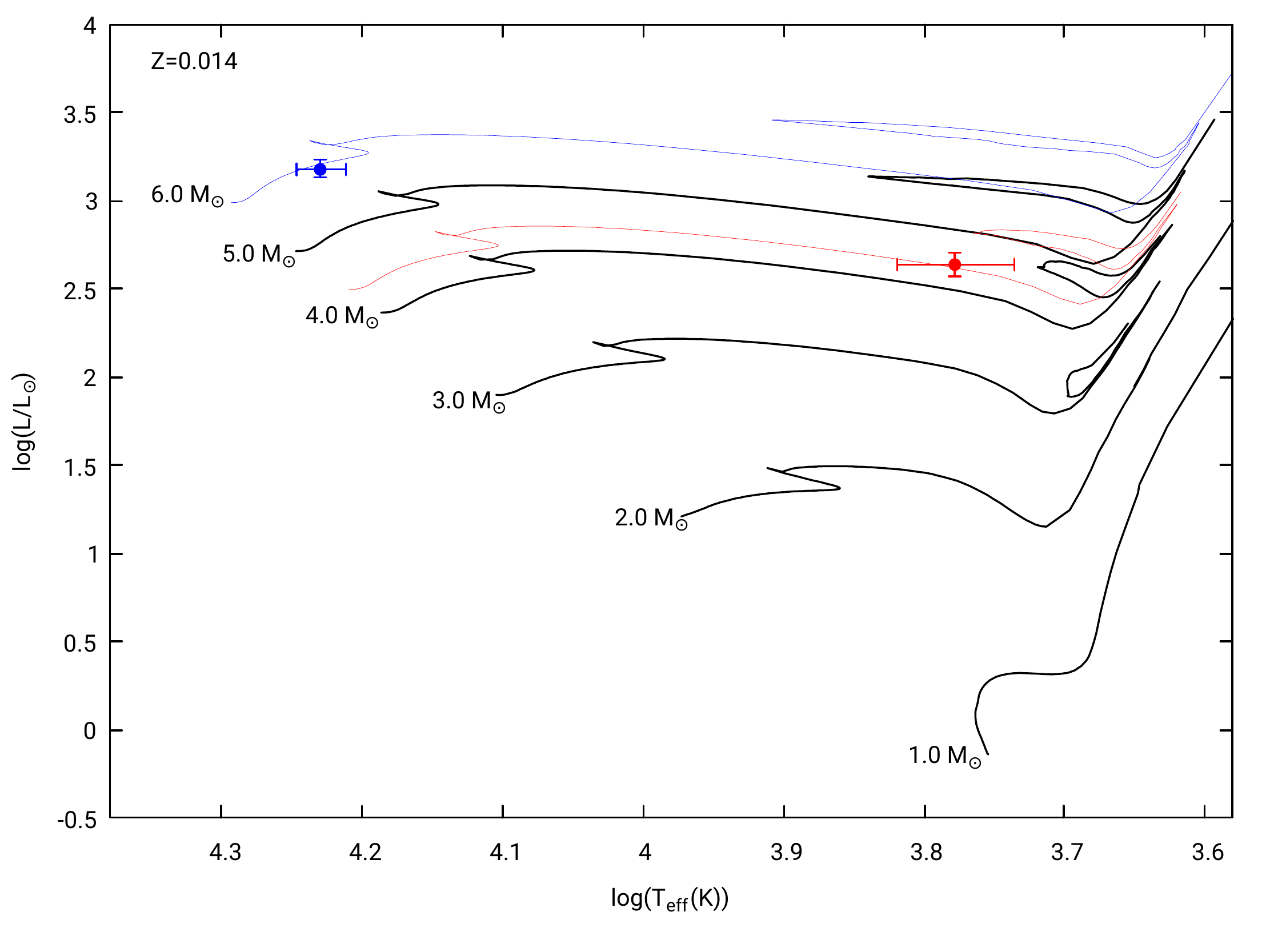}
	\end{center}
	\footnotesize
	{\bf Figure 14.} 
	Hertzsprung-Russell diagram of evolutionary track for non-rotating single stars models (\citet{2012A&A...537A.146E}). The blue line corresponds to $6.0 M_\odot$ while the red line to $4.35 M_{\odot}$. Lines represent stellar tracks without rotation calculated at Z= 0.014.
	\normalsize
\end{figure}

%%%%%%%%%%%%%%%%%%%%%%%%%%%%%%%%%%%%%%%%%%%%%%%%%%%%%%%%%%%%%%%%%%%%%%%%%%%%%%%%%%%%%%%%%%%%%%%%%%%%%%%%%%%%%%%%%%%%%%%%%%%%%%%%%%%%%%%%%
%%%%%%%%%%%%%%%%%%%%%%%%%%%%%%%%%%%%%%%%%%%%%%%%%%%%%%%%%%%%%%%%%%%%%%%%%%%%%%%%%%%%%%%%%%%%%%%%%%%%%%%%%%%%%%%%%%%%%%%%%%%%%%%%%%%%%%%%%
%%%%%%%%%%%%%%%%%%%%%%%%%%%%%%%%%%%%%%%%%%%%%%%%%%%%%%%%%%%%%%%%%%%%%%%%%%%%%%%%%%%%%%%%%%%%%%%%%%%%%%%%%%%%%%%%%%%%%%%%%%%%%%%%%%%%%%%%%

\vskip 0.5cm
{\bf 6.3 Spectral Energy Distribution (SED).}\\

The determination of the physical parameters of astronomical objects from observational data is frequently linked with the use of theoretical models as templates. In order to obtain the broad-band photometric fluxes we compiled all the information available for V495 Cen and built the SEDs with the aid of the Spanish Virtual Observatory SED Analyzer \footnote{http://svo2.cab.inta-csic.es/theory/vosa4/} (\citet{2008A&A...492..277B}). We performed a statistical test to decide which synthetic model best reproduces the observed data. The provided {\textquotedblleft best\textquotedblright} fitting model is the one that minimizes the value of reduced $\chi^{2}$, considering the composite flux as:\\

\begin{equation}
f_{\lambda}=f_{\lambda,0}10^{-0.4 E(B-V)[k(\lambda- V)+R(V)]}
\end{equation}
\\

where\\

\begin{equation}
f_{\lambda,0}=\left(\frac{R_{2}}{d}\right)^{2}\left[\left(\frac{R_{1}}{R_{2}}\right)^{2}f_{1,\lambda}+f_{2,\lambda} \right]
\end{equation}
\\

and  $f_{1}$ and $f_{2}$ are the fluxes of the  star and secondary star respectively, $R(V)\equiv A(\lambda)/E(B-V)$ is the ratio of reddening to extinction at $V$, $d$ is the distance to the system and  $R_{1}/R_{2}$ is the ratio of the primary star radius to the secondary star radius, $k(\lambda-V)\equiv E(\lambda-V)/E(B-V)$ is the normalized extinction curve and was calculated in two steps:\\

\begin{equation}
k=\left \lbrace
\begin{array}{lllll}

\varepsilon \lambda^{-\beta}-R_{v}& if &  {} & x &  < 0.3\\

R_{v}\left(a(x)+\frac{b(x)}{R_{v}}-1\right)& if& 0.3 \leq & x & \leq 8.0 \\

\end{array}
\right.
\end{equation}
\\

\noindent
Where $x\equiv 1/\lambda(\mu m^{-1})$, $\varepsilon=1.19$, $\beta=1.84$ and R$_{v}=3.05$, with formal errors of about $1\%$ (\citet{1990ApJ...357..113M}), however, this value of $\beta$ is not applicable at wavelengths beyond 5 $\mu m$. The parameters a(x) and b(x) are the parametrization coefficients for the extinction law from \citet{1989IAUS..135P...5C}.

The theoretical spectra  were taken from the library of ATLAS9 Kurucz ODFNEW /NOVER models. We used fluxes calculated with metallicity index $[M/H]$= 0.0,  the micro turbulence velocity for both stars was of 0.0 {$\rm{kms^{-1}}$}. We considered the reddening produced by galactic dust as discussed in Section 3.1. 
 The parametric space of the models was reduced to consider:
donor temperature between 4250 and 8000 K with step of 250 K and surface gravity between 0.5 and 3.0 with steps 0.5, gainer temperature between 16000 and 17000 K with step of 1000 K and surface gravity 
between 3.5 and 4.5 with step of 0.5. We fixed the radii and considered all values of the photometry obtained with VOSA from 4280 \AA{} until 220883 \AA.

We implemented a chi-square adjustment considering a distance with a 5$\%$ error, for 7 degrees of freedom, i.e. x$_{critic\, 0.95,7}^{2}$. We found a distance of $d = 2092$ pc with a $\chi^{2}= 1.447$. Due to $\chi^{2} < \chi_{critic\, 0.95,7}^{2}$, then the solution is accepted and the solution obtained in Section 6.1 is consistent with the current results (Fig. 15). 

In addition, we observe a slight infrared excess at wavelengths longer than 10000 \AA{}, which cannot come from the main star itself but from cooler emitting material. This excess is probably formed in the same circumstellar material responsible of the strong H$\alpha$ line emission.

\begin{figure}
	\begin{center}
		\includegraphics[width=8 cm,angle=0]{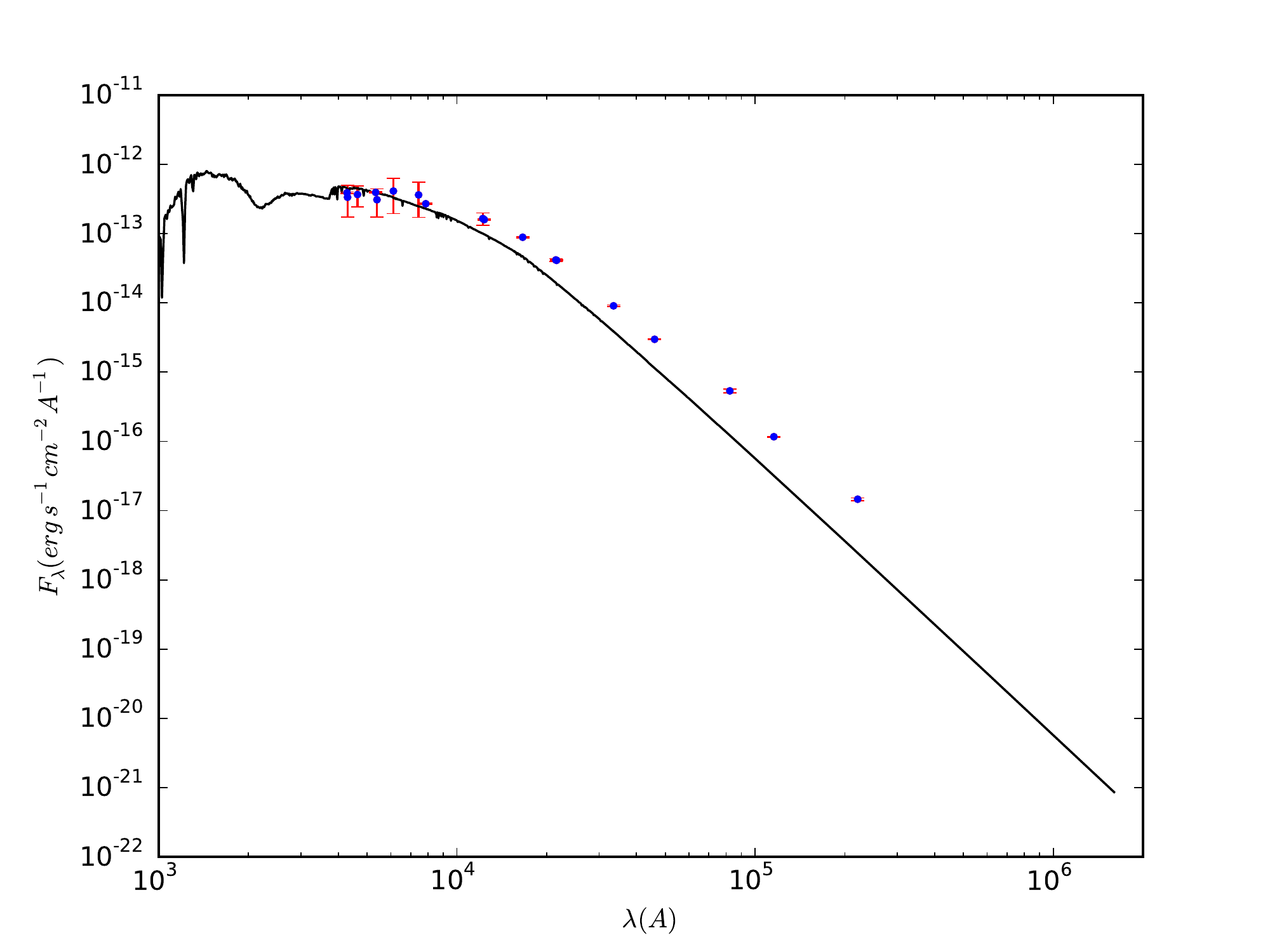}
	\end{center}
	\footnotesize
	{\bf Figure 15.} The best fit to the spectral energy distribution (SED) considers a two-star composite spectrum; it reproduces well the bluer part but exhibits a significant excess from the near-infrared that is probably formed in the circumstellar material. The black and orange theoretical spectrum is given by equation 14.
	\normalsize
\end{figure}
%%%%%%%%%%%%%%%%%%%%%%%%%%%%%%%%%%%%%%%%%%%%%%%%%%%%%%%%%%%%%%%%%%%%%%%%%%%%%%%%%%%%%%%%%%%%%%%%%%%%%%%%%%%%%%%%%%%%%%%%%%%%%%%%%%%%%%%
%%%%%%%%%%%%%%%%%%%%%%%%%%%%%%%%%%%%%%%%%%%%%%%%%%%%%%%%%%%%%%%%%%%%%%%%%%%%%%%%%%%%%%%%%%%%%%%%%%%%%%%%%%%%%%%%%%%%%%%%%%%%%%%%%%%%%%%
%%%%%%%%%%%%%%%%%%%%%%%%%%%%%%%%%%%%%%%%%%%%%%%%%%%%%%%%%%%%%%%%%%%%%%%%%%%%%%%%%%%%%%%%%%%%%%%%%%%%%%%%%%%%%%%%%%%%%%%%%%%%%%%%%%%%%%%
\vskip 0.5cm 

{\bf 7 CONCLUSIONS}\\

In this work we have presented a detailed spectroscopic and photometric analysis of V\,495\,Cen, 
the hitherto longest orbital period Galactic DPV. We conclude:

\begin{itemize}

\item From the photometric study we have determined that V495 shows an orbital period of $P_{o}= 33.492 \pm 0.002$ d and a long period of $P_{l}= 1283$ d. 
%33.481\pm 0.064

\item We have modeled the spectrum and light curve and the best system and orbital parameters we found are given in Table 2. For instance, we determined the mass of the secondary star  $M_{2}= 0.91 \pm 0.2 M_{\astrosun}$, its temperature $T_{2}=6000 \pm 250$ K and radius $R_{2}=19.3 \pm 0.5$, the mass of the primary  $M_{1}=5.76 \pm 0.3 M_{\astrosun}$, its the temperature $T_{1}=16960 \pm 400$K and  radius $R_{1}=4.5 \pm 0.2 R_{\astrosun}$. 

\item The early B type dwarf is surrounded by an optical and geometrically thick  accretion disc of radial extension $R_{d}=40.2\pm 1.3 R_{\astrosun}$. 

\item  At quadratures the donor, gainer and disc contribute 57, 32 and 11\%  to the total flux at the $V$ band, respectively.

\item The best model shows a hot spot located in the outer edge of the disc, 11\% hotter than the surrounding disc and displaced 18.2 degree for the line joining the star centers in the direction of the orbital motion.

\item The spectral energy distribution shows infrared excess indicating the presence of a circumstellar material. 

\item We found a lower limit to the distance  to V495 Cen of $2092$ pc with a $\chi^{2}=1.447$ for 7 degrees of freedom and 95\% of confidence level, i.e. the distance for V495 Cen is 2092 $\pm$ 104.6 pc. 

\end{itemize}

\vskip 0.5cm

{\bf ACKNOWLEDGEMENTS}\\

This investigation is based on observations conduced under CNTAC proposal CN2015A-123. This publication makes use of VOSA, developed under Spanish Virtual Observatory project supported from the Spanish MICINN/MINECO trough grant AyA2008-0256, Ay2011-24052. Also we made use of the SIMBAD database, operated at CDS, Strasbourg, France. This research was funded in part by scholarship from the Faculty  of Physical Sciences and Mathematics of the Universidad de Concepción. J.R. and R.E.M. gratefully acknowledge support from the Chilean BASAL Centro de Excelencia en Astrofísica y Tecnologías Afines (CATA) grant PFB-06/200. J.R. thanks the SOCHIAS grant through Gemini-Conicyt Project 32140015. R.E.M. thanks the grant VRID 216.016.002-1.0. I. A. acknowledges support from Fondo Institucional de Becas FIB-UV and Gemini-Conicyt 32120033. G.D. acknowledges the financial support of the Ministry of Education and Science of the Republic of Serbia through the project 176004 "Stellar physics, also we acknowledge the anonymous referee whose comments helped to improve a first version of this manuscript.\\

%%%%%%%%%%%%%%%%%%%%%%%%%%%%%%%%%%%%%%%%%%%%%%%%%%%%%%%%%%%%%%%%%%%%%%%%%%%%%%%%%%%%%%%%%%%%%%%%%%%%%%%%%%%%%%%%%%%%%%%%%%%%%%%%%%%%%%%
%%%%%%%%%%%%%%%%%%%%%%%%%%%%%%%%%%%%%%%%%%%%%%%%%%%%%%%%%%%%%%%%%%%%%%%%%%%%%%%%%%%%%%%%%%%%%%%%%%%%%%%%%%%%%%%%%%%%%%%%%%%%%%%%%%%%%%%
%%%%%%%%%%%%%%%%%%%%%%%%%%%%%%%%%%%%%%%%%%%%%%%%%%%%%%%%%%%%%%%%%%%%%%%%%%%%%%%%%%%%%%%%%%%%%%%%%%%%%%%%%%%%%%%%%%%%%%%%%%%%%%%%%%%%%%%
%%%%%%%%%%%%%%%%%%%%%%%%%%%%%%%%%%%%%%%%%%%%%%%%%%%%%%%%%%%%%%%%%%%%%%%%%%%%%%%%%%%%%%%%%%%%%%%%%%%%%%%%%%%%%%%%%%%%%%%%%%%%%%%%%%%%%%%

\begin{table*}
	\footnotesize
	\caption{Summary of spectroscopic observation. N is number of spectra. The HJD at mid-exposure for the first spectrum series is given, $\Phi_{o}$ and $\Phi_{l}$ refer to the orbital and long-cycle phase, respectively, and are calculated according to Eq. 1 and Eq. 2. Spectral Resolution R $\sim$	80000 (with image slicer) and R$\sim$ 25000 (fiber mode).}
	\normalsize		
	\[
\resizebox{16cm}{!}{$
	\begin{array}{ccccccccrr}
	\hline
	\textmd{UT-date} & \textmd{Observatory/Telescope} & \textmd{Instrument} & \textmd{N} & \textmd{exptime (s)} & \textmd{HJD} & \textmd{$\Phi_{o}$} & \textmd{$\Phi_{l}$}& \textmd{S/N} & \textmd{R}  \\
	\hline
	2015-02-09 &  \textmd{CTIO/1.5m} &   \textmd{CHIRON} &  1 &  1800 &  2457062.78265893 &  0.2438536557 &  0.6899319243 & 214.11 & 25000 \\
	2015-02-09 &  \textmd{CTIO/1.5m} &   \textmd{CHIRON} &  1 &  1800 &  2457062.80354576 &  0.2444773294 &  0.6899482040 & 133.13 & 25000 \\
	2015-02-12 &  \textmd{CTIO/1.5m} &   \textmd{CHIRON} &  1 &  1800 &  2457065.73864590 &  0.3321184204 &  0.6922358892 & 172.66 & 25000 \\
	2015-02-12 &  \textmd{CTIO/1.5m} &   \textmd{CHIRON} &  1 &  1800 &  2457065.75953170 &  0.3327420633 &  0.6922521681 & 139.00 & 25000 \\
	2015-02-15 &  \textmd{CTIO/1.5m} &   \textmd{CHIRON} &  1 &  1800 &  2457068.81094574 &  0.4238562478 &  0.6946305111 & 114.24 & 25000 \\
	2015-02-15 &  \textmd{CTIO/1.5m} &   \textmd{CHIRON} &  1 &  1800 &  2457068.83183151 &  0.4244798898 &  0.6946467900 & 137.29 & 25000 \\
	2015-02-18 &  \textmd{CTIO/1.5m} &   \textmd{CHIRON} &  1 &  1800 &  2457071.65142447 &  0.5086719758 &  0.6968444462 & 229.02 & 25000 \\
	2015-02-18 &  \textmd{CTIO/1.5m} &   \textmd{CHIRON} &  1 &  1800 &  2457071.67231145 &  0.5092956539 &  0.6968607260 & 107.85 & 25000 \\
	2015-02-21 &  \textmd{CTIO/1.5m} &   \textmd{CHIRON} &  1 &  1800 &  2457074.85684834 &  0.6043848414 &  0.6993428280 & 116.52 & 25000 \\
	2015-02-21 &  \textmd{CTIO/1.5m} &   \textmd{CHIRON} &  1 &  1800 &  2457074.85684834 &  0.6043848414 &  0.6993428280 & 113.69 & 25000 \\
	2015-02-25 &  \textmd{CTIO/1.5m} &   \textmd{CHIRON} &  1 &  1800 &  2457078.74075340 &  0.7203569245 &  0.7023700338 &  79.80 & 80000 \\
	2015-02-25 &  \textmd{CTIO/1.5m} &   \textmd{CHIRON} &  1 &  1800 &  2457078.76179291 &  0.7209851571 &  0.7023864325 &  74.70 & 80000 \\
	2015-02-28 &  \textmd{CTIO/1.5m} &   \textmd{CHIRON} &  1 &  1200 &  2457081.64889093 &  0.8071929212 &  0.7046367038 &  41.35 & 80000 \\
	2015-03-03 &  \textmd{CTIO/1.5m} &   \textmd{CHIRON} &  1 &  1200 &  2457084.80281535 &  0.9013680308 &  0.7070949457 &  54.36 & 80000 \\
	2015-03-06 &  \textmd{CTIO/1.5m} &   \textmd{CHIRON} &  1 &  1200 &  2457087.65572745 &  0.9865550149 &  0.7093185717 &  49.53 & 80000 \\
	2015-03-09 &  \textmd{CTIO/1.5m} &   \textmd{CHIRON} &  1 &  1200 &  2457090.67030988 &  0.0765694201 &  0.7116682072 &  57.18 & 80000 \\
	2015-03-12 &  \textmd{CTIO/1.5m} &   \textmd{CHIRON} &  1 &  1200 &  2457093.71585473 &  0.1675083526 &  0.7140419756 &  60.09 & 80000 \\
	2015-03-13 &  \textmd{CTIO/1.5m} &   \textmd{CHIRON} &  1 &  1200 &  2457094.71578943 &  0.1973660624 &  0.7148213480 &  52.11 & 80000 \\
	2015-03-16 &  \textmd{CTIO/1.5m} &   \textmd{CHIRON} &  1 &  1200 &  2457097.70121347 &  0.2865098080 &  0.7171482568 &  53.50 & 80000 \\
	2015-03-19 &  \textmd{CTIO/1.5m} &   \textmd{CHIRON} &  1 &  1200 &  2457100.64933568 &  0.3745397337 &  0.7194460917 &  63.64 & 80000 \\
	2015-03-22 &  \textmd{CTIO/1.5m} &   \textmd{CHIRON} &  1 &  1200 &  2457103.62769326 &  0.4634724772 &  0.7217674928 &  36.06 & 80000 \\
	2015-04-03 &  \textmd{CTIO/1.5m} &   \textmd{CHIRON} &  1 &  1200 &  2457115.62824705 &  0.8218049283 &  0.7311210032 &  76.57 & 80000 \\
	2015-04-06 &  \textmd{CTIO/1.5m} &   \textmd{CHIRON} &  1 &  1200 &  2457118.57643026 &  0.9098366754 &  0.7334188856 &  62.55 & 80000 \\
	2015-04-09 &  \textmd{CTIO/1.5m} &   \textmd{CHIRON} &  1 &  1200 &  2457121.61299987 &  0.0005076103 &  0.7357856585 &  50.84 & 80000 \\
	2015-04-09 &  \textmd{CTIO/1.5m} &   \textmd{CHIRON} &  1 &  3360 &  2457121.68248558 &  0.0025824300 &  0.7358398173 & 102.58 & 80000 \\
	2015-04-09 &  \textmd{CTIO/1.5m} &   \textmd{CHIRON} &  1 &  3360 &  2457121.72158038 &  0.0037497874 &  0.7358702887 & 101.85 & 80000 \\
	2015-04-10 &  \textmd{CTIO/1.5m} &   \textmd{CHIRON} &  1 &  1200 &  2457122.62723699 &  0.0307923855 &  0.7365761785 &  51.96 & 80000 \\
	2015-04-10 &  \textmd{CTIO/1.5m} &   \textmd{CHIRON} &  1 &  3360 &  2457122.68590166 &  0.0325440926 &  0.7366219031 &  68.10 & 80000 \\
	2015-04-11 &  \textmd{CTIO/1.5m} &   \textmd{CHIRON} &  1 &  1200 &  2457123.60255273 &  0.0599149815 &  0.7373363622 &  60.33 & 80000 \\
	2015-04-13 &  \textmd{CTIO/1.5m} &   \textmd{CHIRON} &  1 &  1200 &  2457125.54243230 &  0.1178391251 &  0.7388483494 &  42.48 & 80000 \\
	2015-04-17 &  \textmd{CTIO/1.5m} &   \textmd{CHIRON} &  1 &  1200 &  2457129.53506219 &  0.2370576945 &  0.7419602979 &  59.30 & 80000 \\
	2015-04-20 &  \textmd{CTIO/1.5m} &   \textmd{CHIRON} &  1 &  1200 &  2457132.67083800 &  0.3306908928 &  0.7444043944 &  57.25 & 80000 \\
	2015-04-23 &  \textmd{CTIO/1.5m} &   \textmd{CHIRON} &  1 &  1200 &  2457135.59115892 &  0.4178906814 &  0.7466805603 &  67.59 & 80000 \\
	2015-04-26 &  \textmd{CTIO/1.5m} &   \textmd{CHIRON} &  1 &  1200 &  2457138.60097040 &  0.5077626277 &  0.7490264773 &  44.16 & 80000 \\
	2015-04-29 &  \textmd{CTIO/1.5m} &   \textmd{CHIRON} &  1 &  1200 &  2457141.63049700 &  0.5982232607 &  0.7513877607 &  45.85 & 80000 \\
	2015-05-03 &  \textmd{CTIO/1.5m} &   \textmd{CHIRON} &  1 &  1200 &  2457145.71165279 &  0.7200851833 &  0.7545687083 &  35.71 & 80000 \\
	2015-05-06 &  \textmd{CTIO/1.5m} &   \textmd{CHIRON} &  1 &  1200 &  2457148.69643838 &  0.8092098650 &  0.7568951195 &  56.81 & 80000 \\
	2015-05-09 &  \textmd{CTIO/1.5m} &   \textmd{CHIRON} &  1 &  1200 &  2457151.80714083 &  0.9020943813 &  0.7593196733 &  55.85 & 80000 \\
	2015-05-10 &  \textmd{CTIO/1.5m} &   \textmd{CHIRON} &  1 &  1200 &  2457152.61159559 &  0.9261151266 &  0.7599466840 &  38.55 & 80000 \\
	2015-05-13 &  \textmd{CTIO/1.5m} &   \textmd{CHIRON} &  1 &  1200 &  2457155.58329222 &  0.0148489764 &  0.7622628934 &  45.06 & 80000 \\
	\hline
	\end{array}
$}
	\]
\end{table*}

\begin{table*}
	\footnotesize
	\caption{Radial velocities of the donor and their errors.}
	\normalsize
	\[
	\begin{array}{crc}
	\hline
	\textmd{HJD} & \textmd{RV (km $s^{-1}$)} & \textmd{error (km $s^{-1}$)} \\
	\hline
	2457078.74075340 & -107.007 & 0.464 \\
	2457078.76179291 & -106.801 & 0.391 \\
	2457081.64889093 & -98.052  & 0.496 \\
	2457084.80281535 & -56.000  & 0.649 \\
	2457087.65572745 & -2.660   & 0.705 \\
	2457090.67030988 & 54.155   & 0.854 \\
	2457093.71585473 & 94.720   & 1.105 \\
	2457094.71578943 & 101.841  & 0.977 \\
	2457097.70121347 & 101.745  & 1.098 \\
	2457100.64933568 & 71.432   & 1.260 \\
	2457103.62769326 & 15.153   & 1.791 \\
	2457115.62824705 & -93.987  & 0.728 \\
	2457118.57643026 & -50.599  & 0.944 \\
	2457121.61299987 & 7.012    & 0.720 \\
	2457121.68248558 & 8.738    & 0.769 \\
	2457121.72158038 & 9.406    & 0.749 \\
	2457122.62723699 & 26.157   & 1.172 \\
	2457122.68590166 & 27.430   & 1.147 \\
	2457123.60255273 & 44.459   & 1.214 \\
	2457125.54243230 & 75.944   & 1.018 \\
	2457129.53506219 & 105.475  & 1.007 \\
	2457132.67083800 & 89.557   & 0.901 \\
	2457135.59115892 & 48.835   & 1.292 \\
	2457138.60097040 & -18.296  & 1.658 \\
	2457141.63049700 & -69.054  & 1.008 \\
	2457145.71165279 & -107.108 & 0.664 \\
	2457148.69643838 & -97.126  & 0.781 \\
	2457151.80714083 & -54.372  & 1.042 \\
	2457152.61159559 & -41.587  & 1.021 \\
	2457155.58329222 & 15.412   & 0.771 \\
	\hline
	\end{array}
	\]
\end{table*}

\begin{table*}
	\footnotesize
	\caption{Radial velocities of the gainer and their errors.}
	\normalsize	
	\[
	\begin{array}{crc}
	\hline
	\textmd{HJD} & \textmd{RV (km $s^{-1}$)} & \textmd{error (km $s^{-1}$)} \\
	\hline
	2457121.61299987 & 16.455  & 0.611 \\
	2457121.68248558 & 17.276  & 0.599 \\
	2457121.72158038 & 11.927  & 0.600 \\
	2457123.60255273 & 7.559   & 0.624 \\
	2457090.67030988 & 5.428   & 0.624 \\
	2457093.71585473 & -4.516  & 0.610 \\
	2457094.71578943 & -4.310   & 0.608 \\
	2457129.53506219 & -4.208  & 0.608 \\
	2457062.78265893 & -6.731  & 0.603 \\
	2457062.80354576 & -6.808  & 0.603 \\
	2457097.70121347 & -11.708 & 0.609 \\
	2457132.67083800 & -15.873 & 0.604  \\
	2457065.73864590 & -16.498 & 0.611 \\
	2457065.75953170 & -15.399 & 0.611 \\
	2457138.60097040 & -16.877 & 0.619 \\
	2457071.65142447 & -12.784 & 0.627 \\
	2457071.67231145 & -14.066 & 0.616 \\
	2457074.85684834 & -7.853  & 0.614 \\
	2457145.71165279 & 8.475   & 0.614 \\
	2457078.74075340 & 9.035   & 0.606 \\
	2457078.76179291 & 8.733   & 0.605 \\
	2457148.69643838 & 15.724  & 0.607 \\
	2457087.65572745 & 8.391   & 0.616 \\
	\hline
	\end{array}
	\]
\end{table*}

\begin{table*}
	\footnotesize
	\caption{Radial velocities of He\,I\,5875 and their errors}
	\normalsize	
	\[
	\begin{array}{crrr}
	\hline
	\textmd{HJD} & \textmd{RV (km $s^{-1}$)} & \textmd{error (km $s^{-1}$)} \\
	\hline
	2457121.61299987 & -130.882 & 0.685 \\
	2457121.68248558 & -134.489 & 0.607 \\
	2457121.72158038 & -128.763 & 0.363 \\
	2457155.58329222 & -122.714 & 0.032 \\
	2457122.62723699 & -110.354 & 3.539 \\
	2457122.68590166 & -104.390 & 0.660 \\
	2457123.60255273 & -82.371 & 0.255 \\
	2457090.67030988 & -80.060 & 0.291 \\
	2457125.54243230 & -62.002 & 1.056 \\
	2457093.71585473 & -37.893 & 1.700 \\
	2457094.71578943 & -38.739 & 2.546 \\
	2457129.53506219 & -33.361 & 0.696 \\
	2457062.78265893 & -26.676 & 0.435 \\
	2457062.80354576 & -27.353 & 2.761 \\
	2457097.70121347 & -42.736 & 0.463 \\
	2457065.73864590 & -40.997 & 1.276 \\
	2457065.75953170 & -56.653 & 1.059 \\
	2457100.64933568 & -58.969 & 0.314 \\
	2457135.59115892 & -82.542 & 1.205 \\
	2457068.81094574 & -93.843 & 7.426 \\
	2457068.83183151 & -99.579 & 1.821 \\
	2457103.62769326 & -116.214 & 0.815 \\
	2457138.60097040 & -138.043 & 1.183 \\
	2457071.65142447 & -139.524 & 1.116 \\
	2457074.85684834 & -211.019 & 0.267 \\
	2457074.85684834 & -211.019 & 0.267 \\
	2457145.71165279 & -245.749 & 0.820 \\
	2457078.74075340 & -244.460 & 0.849 \\
	2457078.76179291 & -239.287 & 2.241 \\
	2457081.64889093 & -232.824 & 1.144 \\
	2457148.69643838 & -234.363 & 0.394 \\
	2457084.80281535 & -180.051 & 1.567 \\
	2457151.80714083 & -189.062 & 0.497 \\
	2457118.57643026 & -172.555 & 2.148 \\
	2457152.61159559 & -179.926 & 1.442 \\
	2457087.65572745 & -140.905 & 1.524 \\
	\hline
	\end{array}
	\]
\end{table*}

\begin{table*}
	\footnotesize
	\caption{Photometry points extracted from Vizier Catalogue with a search radius of 2 arsec to build the Spectral Energy Distribution}
	\normalsize
	\[
	\resizebox{14cm}{!}{$
	\begin{array}{l c c l}
	\hline
	\noalign{\smallskip}
	\textrm {wavelength}   	& \textrm{Obs. Flux} 		& \textrm{Obs. Error}& \textrm{Filter ID}\\
	\textrm{~~~~A}   		& \textrm{erg/s/cm$^{2}$/A}	& \textrm{erg/s/cm$^{2}$/A}\\
	\hline
	\hline
	4280                &3.8961272708038E-13    &1.0765397488993E-14& \textrm{TYCHO/TYCHO.B}\\ 
	4297.1691826842     &3.3568742904622E-13    &1.6077336496569E-13& \textrm{Misc/APASS.B}\\
	4640.4198339016     &3.6790221791186E-13    &1.2469697114109E-13& \textrm{Misc/APASS.sdss$_{g}$}\\  
	5340                &3.9470079496967E-13    &9.0883216669006E-15& \textrm{TYCHO/TYCHO.V}\\  
	5394.2913582927     &3.0914068920168E-13    &1.3638523748095E-13& \textrm{Misc/APASS.V}\\  
	6122.3296283695     &4.1249010772103E-13    &2.1731276951292E-13& \textrm{Misc/APASS.sdss$_{r}$}\\  
	7439.4904529922     &3.6321232166977E-13    &1.9135168108482E-13& \textrm{Misc/APASS.sdss$_{i}$}\\  
	7862.1015966        &2.7060061005582E-13    &4.9846474469570E-15& \textrm{DENIS/DENIS.I}\\  
	12210.6027581       &1.6530536695297E-13    &3.3495411288929E-14& \textrm{DENIS/DENIS.J}\\  
	12350               &1.5958995899691E-13    &3.3807190373159E-15& \textrm{2MASS/2MASS.J}\\  
	16620               &8.8762578525412E-14    &1.6350671210266E-15& \textrm{2MASS/2MASS.H}\\  
	21465.009653        &4.1737018974359E-14    &1.9220607543274E-15& \textrm{DENIS/DENIS.Ks}\\  
	21590               &4.1204825037746E-14    &9.1082511255687E-16& \textrm{2MASS/2MASS.Ks}\\  
	33526               &9.0757866535675E-15    &2.7585057793523E-16& \textrm{WISE/WISE.W1}\\  
	46028               &2.9820652029785E-15    &5.2185087508631E-17& \textrm{WISE/WISE.W2}\\  
	82283.5545614       &5.3621475465995E-16    &3.7105529678368E-17& \textrm{AKARI/IRC.S9W}\\  
	115608              &1.1671349817430E-16    &1.6124565662840E-18& \textrm{WISE/WISE.W3}\\  
	220883              &1.4612560240564E-17    &6.4601593690368E-19& \textrm{WISE/WISE.W4}\\  
	\hline
	\end{array}
	$}
	\]
\end{table*}

\end{document}